\documentclass[11pt,a4paper]{article}
\usepackage{cite}
\usepackage{graphicx}
\usepackage{amssymb}
\usepackage{amsmath}
\usepackage{amsfonts}
\usepackage{dsfont}
\usepackage{mathtools}
\usepackage{array}
\usepackage{rotating}
\usepackage{bbold,amsfonts}

\usepackage[utf8]{inputenc}
\usepackage{bm}
\usepackage{xcolor}
\usepackage{braket}

\usepackage[height=8.8in,width=6.45in]{geometry}
\usepackage[font=small,labelfont=bf]{caption}
\usepackage[hidelinks]{hyperref}
\usepackage{booktabs,float,slashed}
\usepackage{standalone}
\usepackage{caption}
\usepackage{subcaption}
\numberwithin{equation}{section}

\newcommand{\beq}{\begin{equation}}
\newcommand{\eeq}{\end{equation}}

\newcommand{\overbar}[1]{\mkern 1.5mu\overline{\mkern-1.5mu#1\mkern-1.5mu}\mkern 1.5mu}

\newcommand{\bcP}{\boldsymbol{\mathcal{P}}}

\DeclareMathOperator{\tr}{tr}

\newcommand{\ii}{\mathrm{i}}

\makeatletter
\newcommand*{\letterdef@}{}
\newcommand*{\letterdef}[3]{%
	\def\letterdef@##1{\expandafter\newcommand\csname #1\endcsname{#2{##1}}}%
	\@tfor\@tempa :=#3\do{\expandafter\letterdef@\expandafter{\@tempa}}}
\makeatother
\letterdef{c#1} {\mathcal}{ABCDEFGHIJKLMNOPQRSTUVWXYZ} 
\letterdef{rm#1}{\mathrm} {dDeimM} 

\newcommand{\Xx}{\mathsf{X}}
\newcommand{\Dx}{\mathsf{D}}
\newcommand{\Sx}{\mathsf{S}}
\newcommand{\Yx}{\mathsf{Y}}

\newdimen\tableauside\tableauside=1.0ex
\newdimen\tableaurule\tableaurule=0.4pt
\newdimen\tableaustep
\def\phantomhrule#1{\hbox{\vbox to0pt{\hrule height\tableaurule
			width#1\vss}}}
\def\phantomvrule#1{\vbox{\hbox to0pt{\vrule width\tableaurule
			height#1\hss}}}
\def\sqr{\vbox{%
		\phantomhrule\tableaustep
		\hbox{\phantomvrule\tableaustep\kern\tableaustep\phantomvrule\tableaustep}%
		\hbox{\vbox{\phantomhrule\tableauside}\kern-\tableaurule}}}
\def\squares#1{\hbox{\count0=#1\noindent\loop\sqr
		\advance\count0 by-1 \ifnum\count0>0\repeat}}
\def\tableau#1{\vcenter{\offinterlineskip
		\tableaustep=\tableauside\advance\tableaustep by-\tableaurule
		\kern\normallineskip\hbox
		{\kern\normallineskip\vbox
			{\gettableau#1 0 }%
			\kern\normallineskip\kern\tableaurule}%
		\kern\normallineskip\kern\tableaurule}}
\def\gettableau#1 {\ifnum#1=0\let\next=\null\else
	\squares{#1}\let\next=\gettableau\fi\next}
\allowdisplaybreaks

\begin{document}
\begin{titlepage}
\vspace*{10mm}
\begin{center}
{\LARGE \bf 
	 	Strong-coupling results for $\cN=2$ superconformal quivers and holography
}

\vspace*{15mm}

{\Large M. Bill\`o${}^{\,a,d}$, M. Frau${}^{\,a,d}$, F. Galvagno${}^{\,b}$, A. Lerda${}^{\,c,d}$, A. Pini${}^{\,d}$}

\vspace*{8mm}
	
${}^a$ Universit\`a di Torino, Dipartimento di Fisica,\\
			Via P. Giuria 1, I-10125 Torino, Italy
			\vskip 0.3cm
			
${}^b$ Institut f\"ur Theoretische Physik, ETH Z\"urich\\
		Wolfgang-Pauli-Strasse 27, 8093 Z\"urich, Switzerland			
			\vskip 0.3cm
			
${}^c$  Universit\`a del Piemonte Orientale,\\
			Dipartimento di Scienze e Innovazione Tecnologica\\
			Viale T. Michel 11, I-15121 Alessandria, Italy
			\vskip 0.3cm
			
${}^d$   I.N.F.N. - sezione di Torino,\\
			Via P. Giuria 1, I-10125 Torino, Italy

\vskip 0.8cm
	{\small
		E-mail:
		\texttt{billo,frau,lerda,apini@to.infn.it,fgalvagno@phys.ethz.ch}
	}
\vspace*{0.8cm}
\end{center}

\begin{abstract}
We consider $\cN=2$ superconformal quiver gauge theories in four dimensions and 
evaluate the chiral/anti-chiral correlators of single-trace operators. We show that 
it is convenient to form particular twisted and untwisted combinations of these operators 
suggested by the dual holographic description of the theory. The various twisted sectors 
are orthogonal and the correlators in each sector have always the same structure, as we 
show at the lowest orders in perturbation theory with Feynman diagrams. Using 
localization we then map the computation to a matrix model. In this way we are able to 
obtain formal expressions for the twisted correlators in the planar limit that are valid for 
all values of the 't Hooft coupling $\lambda$, and find that they are proportional to 
$1/\lambda$ at strong coupling. We successfully test the correctness of our extrapolation 
against a direct numerical evaluation of the matrix model and argue that the 
$1/\lambda$ behavior qualitatively agrees with the holographic description.
\end{abstract}
\end{titlepage}
\setcounter{tocdepth}{2}
\tableofcontents
\vspace{1cm}
\section{Introduction}
\label{sec:intro}
Four-dimensional gauge theories with a high amount of supersymmetry, such as 
$\cN=4$ and $\cN=2$ super Yang-Mills (SYM) theories, represent a formidable 
playground in the quest for exact results at the quantum level and for ways to control the 
strong-coupling regime.
Among the powerful tools at our disposal to study these theories are the use of 
localization techniques \cite{Pestun:2016jze} and, in some cases, the use of a 
holographic description \cite{Aharony:1999ti}.  

Localization allows one to map the partition function and other protected observables of 
a supersymmetric gauge theory defined on a four-sphere to quantities in a matrix model 
\cite{Pestun:2007rz}; when the quantum theory is conformal, this matrix model captures 
also the corresponding observables in flat space. While the matrix model associated to 
the $\cN=4$ SYM theory is Gaussian, the one for $\cN=2$ theories has a complicated 
potential with contributions 
at any order in the gauge coupling constant $g$. Nevertheless, it can be efficiently used 
to study a whole set of observables, like for instance the Wilson loop vacuum expectation 
value \cite{Andree:2010na,Rey:2010ry,Passerini:2011fe,Bourgine:2011ie}, 
the chiral/anti-chiral correlators \cite{Baggio:2014sna,Gerchkovitz:2016gxx,Baggio:2016skg,Rodriguez-Gomez:2016cem,Rodriguez-Gomez:2016ijh,Billo:2017glv,Billo:2019job,Billo:2019fbi,Fiol:2021icm}, the correlators of chiral 
operators and Wilson loops \cite{Semenoff:2001xp,Billo:2018oog}, as well as the 
Bremsstrahlung function \cite{Fiol:2015spa,Fiol:2015mrp,Bianchi:2018zpb,Bianchi:2019dlw}.

Even if most of these results have been obtained at weak coupling, it is obviously 
interesting to extend them also at strong coupling. Within the matrix model approach, 
some important progress in this direction has been achieved in the large-$N$ limit, where $N$ is the number of colors, keeping the 't Hooft coupling $\lambda = N\,g^2$ 
fixed\,%
\footnote{In the 't Hooft limit, the instanton contributions are suppressed. Of course it 
would be very interesting to consider also the large-$N$ limit at fixed $g$ in which case 
the instantons must be taken into account.}. However, if the supersymmetry is not 
maximal, finding precise and explicit results when $\lambda$ is large, is not trivial.

In this work we focus on a particular class of theories which are ``very close'' to 
the $\cN=4$ SYM theory, namely the $\cN=2$ superconformal quiver theories with 
gauge group $\mathrm{SU}(N)^M$ and bi-fundamental matter. These theories, which can 
be obtained from D3-branes in Type II B string theory on a $\mathbb{Z}_M$ orbifold 
background \cite{Douglas:1996sw,Kachru:1998ys,Oz:1998hr,Gukov:1998kk,Park:1998zh}, have been extensively studied in integrability contexts \cite{Gadde:2009dj,Gadde:2010zi,Pomoni:2011jj,Gadde:2012rv,Pomoni:2013poa,Mitev:2014yba,Mitev:2015oty,Pomoni:2019oib,Niarchos:2019onf,Niarchos:2020nxk,Pomoni:2021pbj} as well as using localization \cite{Pini:2017ouj,Fiol:2020ojn,Zarembo:2020tpf,Ouyang:2020hwd,Galvagno:2020cgq,Beccaria:2021ksw,Galvagno:2021bbj}, but mostly at the perturbative level.
A related class of models is represented by the $\cN=2$ theories on orientifolds (see for 
example \cite{Park:1998zh}). Recently, in the specific instance of orientifold theory 
with gauge group SU($N$) and matter in the symmetric and anti-symmetric representations, 
which was dubbed $\mathbf{E}$ theory in \cite{Billo:2019fbi}, 
significant progress has been made in extracting exact results from the localization matrix model 
in the large-$N$ limit and in exploring the strong-coupling regime \cite{Beccaria:2020hgy,Beccaria:2021vuc,Beccaria:2021hvt,Beccaria:2021ism}. Here we will extend this analysis to the $\cN=2$
quiver theories. 

One key advantage of dealing with the quiver theories is that, being obtained from the 
$\cN=4$ SYM theory by means of a simple orbifold projection, they inherit from it a 
holographic dual, namely the near-horizon limit of the Type II B string theory in presence 
of D3-branes on a $\mathbb{Z}_M$ orbifold background \cite{Kachru:1998ys,Oz:1998hr,Gukov:1998kk}. The holographic description of a conformal theory in four 
dimensions has its paradigm in the representation of the $\cN=4$ SYM theory as Type II B string theory on the $\mathrm{AdS}_5\times S^5$ background \cite{Maldacena:1997re}. In the large-$N$ limit and for large values of the 't Hooft coupling, both string 
loop corrections and world-sheet corrections are suppressed, so that the theory reduces 
to Type II B supergravity on $\mathrm{AdS}_5\times S^5$. 
In the case of $\cN=2$ superconformal quiver theories with gauge group $\mathrm{SU}(N)^M$, the holographic dual organizes itself in one untwisted and $(M-1)$ twisted sectors. In the near-horizon limit, the untwisted sector corresponds to the $\mathbb{Z}_M$-invariant part of the $\mathrm{AdS}_5\times S^5$ theory, while the twisted sectors are described by a six-dimensional supergravity model on $\mathrm{AdS}_5\times S^1$ \cite{Gukov:1998kk}. 

In the weak-coupling regime, the constituent fields of the quiver theories 
can be given a simple interpretation in terms of open strings attached to fractional D3-
branes in the orbifold background \cite{Douglas:1996sw}; in fact the adjoint fields arise 
from open strings starting and ending on the same fractional brane, while the bi-
fundamental matter field correspond to open string stretching between two different 
branes. This open-string description has a closed-string counterpart in terms of 
boundary states (for a review see for instance \cite{Bertolini:2001gq}). As discussed in 
\cite{Billo:2000yb}, the consistent boundary states corresponding to the fractional branes 
of the $\mathbb{Z}_M$ orbifold are simple combinations, dictated by the Cardy formula, 
of the so-called Ishibashi states in the various (un)twisted closed string sectors. Here we 
take inspiration from this fact and do not work directly with the single-trace gauge 
invariant chiral operators defined in each node. Rather we consider (un)twisted 
combinations which are precisely constructed according to the Cardy formula. 

This change of basis in the space of chiral operators makes the computation of the 
twisted correlators much more transparent. We show this by first working directly with 
the Feynman diagrams of the gauge theory in the large-$N$ limit and at the lowest 
perturbative orders, where the correlators factorize in the various sectors. Indeed, within 
each sector, the perturbative expansion has always the same structure, the difference 
between the various sectors being captured by a single numerical factor. To proceed 
further, we then resort to the matrix model description, and also in this context we 
introduce (un)twisted combinations of the matrix operators defined in each node of the 
quiver. At tree level the multiple correlators of such operators factorize in the various 
sectors and their expressions become
very simple in the planar limit and are captured by an effective Wick rule. This allows us 
to obtain closed-form expressions for the correlators also when the effects of the 
interaction terms in the matrix model are included. In this way we are able to show that 
in the large-$N$ limit the untwisted correlators do not depart from the corresponding 
ones in the $\cN=4$ SYM theory, while the twisted ones can be written in terms of an 
infinite matrix which is the same in all sectors and whose
elements are given by integrals of products of Bessel functions.

This closed-form expression, which is one of the main results of this paper, contains the 
entire dependence on the 't Hooft coupling of the twisted correlators and thus can be 
used to explore them in the various regimes of the theory.
At weak coupling, we can expand the closed-form formula in powers 
of $\lambda$ and push the computation to any desired perturbative order without any 
difficulty. The resulting series, whose lowest terms perfectly agree with the Feynman 
diagram calculations,
have a convergence radius $\lambda = \pi^2$, but they can be resummed à la Padé to 
obtain expressions that remain stable well beyond this limit. If $\lambda$ is large, we 
can instead exploit 
the properties of the Bessel functions to obtain the leading term of the twisted correlators
in the asymptotic expansion at strong coupling, which turns out to be proportional to $1/\lambda$
if one normalizes with respect to the $\cN=4$ correlators. 
We successfully test our matrix model results for the twisted correlators, namely the Padé-resummed perturbative expansions and the leading strong-coupling behavior, comparing them to a direct numerical evaluation of the matrix integral by means of Monte Carlo methods. Of course, our Monte Carlo simulation is carried out at finite $N$ and without instantons, but when we increase the value of $N$ we find that the agreement becomes better and better. This check represents a clear indication of the validity of our manipulations.

The strong-coupling behavior of the correlators should be accessible also from the holographic perspective. We have already mentioned that it is the holographic prescription that suggests to consider the particular (un)twisted combinations of chiral operators in the first place. Moreover, the spectrum of the $\mathrm{AdS}_5\times S^1$ supergravity \cite{Gukov:1998kk} does indeed contain scalar fields of squared mass $n(n-4)$ in one-to-one correspondence with the twisted operators of dimension $n$ of the quiver gauge theory. Correlators of the latter should then be captured, according to the general AdS/CFT philosophy, by the value of effective action for these fields with prescribed boundary conditions. However, if we only consider the 2-point functions, there is a normalization ambiguity. Despite this, we find interesting that the twisted effective supergravity action is quadratic and that its {relative} overall normalization with respect to the untwisted Type II B supergravity action is proportional to $1/\lambda$. These two features are in qualitative agreement with our results obtained from the localization approach.           

The paper is organized as follows: in Section~\ref{sec:model} we define the quiver theory, introduce the (un)twisted operators and present their description in terms of the holographic dual supergravity
fields; in Section~\ref{secn:diagrams} we sketch the calculation of the first perturbative terms in the
(un)twisted correlators using Feynman diagrams in the planar limit; 
in Section~\ref{sec:matrix} we introduce the matrix
model and discuss the properties of the (un)twisted correlators in the large-$N$ limit; in Section~\ref{sec:exact} we derive the exact closed-form expression for the twisted correlators, discuss their
leading strong-coupling term and report the results of the numerical evaluations with Monte Carlo methods; in Section~\ref{sec:discussion} we present our conclusion and discuss how our strong-coupling extrapolation qualitatively agrees with the holographic picture. More technical details are
collected in the appendices, where we also show how the orientifold 
$\mathbf{E}$ theory is obtained from the 2-node quiver model with a suitable projection.

\section{The quiver gauge theory}
\label{sec:model}
\begin{figure}[ht]
\begin{center}
\includegraphics[width=0.42\textwidth]{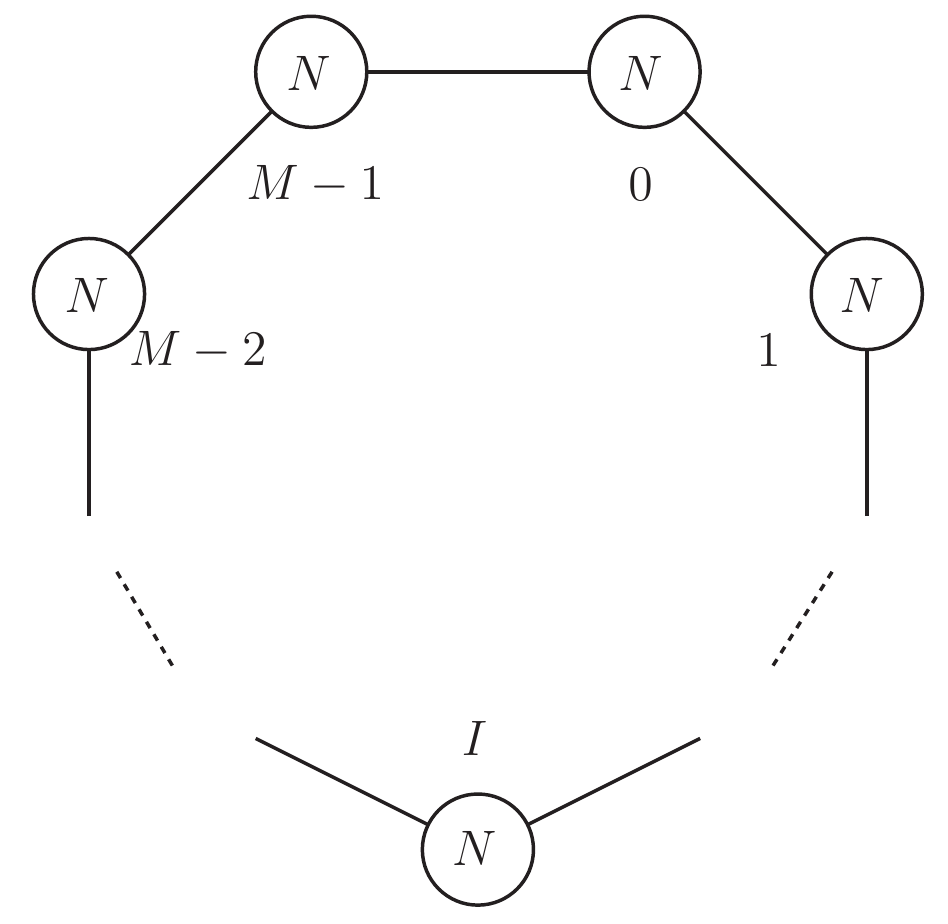}
\caption{The quiver diagram representing the $\mathcal{N}=2$ superconformal
theory with gauge group 
$\mathrm{SU}(N)\times \mathrm{SU}(N) \times \ldots\times \mathrm{SU}(N)$ 
and matter multiplets in the bi-fundamental representation.
\label{fig:quiver}}
\end{center}
\end{figure}

We consider the $\cN=2$ quiver theory described by the diagram 
in Fig.~\ref{fig:quiver}, where each node represents a gauge group factor SU($N$) and each line represents a hypermultiplet in the bi-fundamental representation 
$(\mathbf{N},\overbar{\mathbf{N}})$. 
\noindent
In total there are $M$ nodes (labeled by an index $I=0,\ldots, M-1$), $M$ adjoint
vector multiplets and $M$ bi-fundamental hypermultiplets. We take all gauge coupling
constants to be equal in all nodes and call them $g$. This quiver theory is superconformal since there are $2N$ fundamental flavors at each SU($N$) node, and thus
the coupling constant $g$ does not run.

Denoting the complex scalar field of the $I$-th vector multiplet 
by $\Phi_I(\vec{x})$, we construct the single-trace chiral and anti-chiral operators
\begin{equation}
O_{n}^{(I)}(\vec{x})=
\tr \Phi_I(\vec{x})^n\qquad\mbox{and}\qquad
\overbar{O}_{n}^{(I)}(\vec{x})=
\tr \Phi_I^\dagger(\vec{x})^n
\label{OIn}
\end{equation}
for $n\geq 2$, and re-organize them in the following combinations
\begin{subequations}
\begin{align}
U_n(\vec{x})&=\frac{1}{\sqrt{M}}\Big(O_{n}^{(0)}(\vec{x})+O_{n}^{(1)}(\vec{x})+
\ldots+O_{n}^{(M-1)}(\vec{x})\Big)~,\label{Un}\\
T_{\alpha,n}(\vec{x})&=\frac{1}{\sqrt{M}}\sum_{I=0}^{M-1} \rho^{-\alpha I}\,
O_{n}^{(I)}(\vec{x})~,\label{Tnalpha}
\end{align}
\label{UT}%
\end{subequations}
where $\alpha=1,\ldots,M-1$ and $\rho$ is the $M$-th root of unity:
\begin{equation}
\rho=\rme^{\frac{2\pi\ii}{M}}~.
\label{rho}
\end{equation}
For reasons that will be clear in the following, the operators $U_n$ are called ``untwisted'', while the
operators $T_{\alpha,n}$ are called ``twisted''.
The anti-chiral operators $\overbar{U}_n$ and $\overbar{T}_{\alpha,n}$ are defined in a similar way with $O_{n}^{(I)}$ replaced by $\overbar{O}_{n}^{(I)}$. All these operators are conformal primary operators with dimension $n$.

For example, in the $M=3$ quiver theory we have
\begin{align}
U_n(\vec{x})&=\frac{1}{\sqrt{3}}\Big(O_{n}^{(0)}(\vec{x})+O_{n}^{(1)}(\vec{x})+O_{n}^{(2)}(\vec{x})\Big)~,\notag\\
T_{1,n}(\vec{x})&=\frac{1}{\sqrt{3}}\Big(O_{n}^{(0)}(\vec{x})+\rho^2\,
O_{n}^{(1)}(\vec{x})+\rho\,O_{n}^{(2)}(\vec{x})\Big)~,\\
T_{2,n}(\vec{x})&=\frac{1}{\sqrt{3}}\Big(O_{n}^{(0)}(\vec{x})+\rho\,
O_{n}^{(1)}(\vec{x})+\rho^2\,O_{n}^{(2)}(\vec{x})\Big)~,\notag
\end{align}
and
\begin{align}
\overbar{U}_n(\vec{x})&=\frac{1}{\sqrt{3}}\Big(\overbar{O}_{n}^{(0)}(\vec{x})+\overbar{O}_{n}^{(1)}(\vec{x})+\overbar{O}_{n}^{(2)}(\vec{x})\Big)~,\notag\\
\overbar{T}_{1,n}(\vec{x})&=\frac{1}{\sqrt{3}}\Big(\overbar{O}_{n}^{(0)}(\vec{x})+\rho^2\,
\overbar{O}_{n}^{(1)}(\vec{x})+\rho\,\overbar{O}_{n}^{(2)}(\vec{x})\Big)~,\\
\overbar{T}_{2,n}(\vec{x})&=\frac{1}{\sqrt{3}}\Big(\overbar{O}_{n}^{(0)}(\vec{x})+\rho\,
\overbar{O}_{n}^{(1)}(\vec{x})+\rho^2\,\overbar{O}_{n}^{(2)}(\vec{x})\Big)~,\notag
\end{align}
with $\rho=\rme^{\frac{2\pi\ii}{3}}$. Notice that $\overbar{U}_n=U_n^{\,\dagger}$, 
$\overbar{T}_{1,n}=T_{2,n}^{\,\dagger}$ and $\overbar{T}_{2,n}=T_{1,n}^{\,\dagger}$. 
For a generic $M$,
these complex conjugations become
\begin{equation}
\overbar{U}_n=U_n^{\,\dagger}\quad\mbox{and}\quad\overbar{T}_{\alpha,n}=T_{M-\alpha,n}^{\,\dagger}~.
\label{conjugation}
\end{equation}

Our goal is to compute the 2-point functions of these operators and study them in the 
large-$N$ limit. Of course, only the 2-point functions between chiral and anti-chiral operators are non-vanishing. Furthermore, it is not difficult to realize that the twisted and untwisted operators are mutually orthogonal, namely
\begin{equation}
\big\langle U_n^{\phantom{\dagger}}(\vec{x})\,T_{\alpha,m}^{\,\dagger}(\vec{0})\big\rangle=
\big\langle T_{\alpha,n}^{\phantom{\dagger}}(\vec{x})\,{U}_m^{\,\dagger}(\vec{0})\big\rangle=0
\label{mixedUT}
\end{equation}
for any $\alpha$, $n$ and $m$. Thus, we are left to compute the 2-point functions of two untwisted
and two twisted operators, which take the form
\begin{subequations}
\begin{align}
\big\langle U_n^{\phantom{\dagger}}(\vec{x})\,{U}_{m}^{\,\dagger}(\vec{0})\big\rangle&=\frac{G_n\phantom{\big|}}{\big(
4\pi^2\vec{x}^{\,2}\big)^n\phantom{\Big|}}\,\delta_{n,m}~,\label{UU}\\[1mm]
\big\langle T_{\alpha,n}^{\phantom{\dagger}}(\vec{x})\,{T}_{\beta,m}^{\,\dagger}(\vec{0})\big\rangle&=\frac{G_{\alpha,n}\phantom{\big|}}{\big(
4\pi^2\vec{x}^{\,2}\big)^n\phantom{\Big|}}\,\delta_{n,m}\,\delta_{\alpha,\beta}~,
\label{TT}
\end{align}
\label{UUTT}%
\end{subequations}
where $G_n$ and $G_{\alpha,n}$ are non-trivial functions of the gauge coupling $g$
and of $N$. A few specific examples of these 2-point functions 
have been considered at large $N$ in \cite{Pini:2017ouj} and at finite $N$ in \cite{Niarchos:2020nxk}, while a more systematic study has been recently presented in \cite{Galvagno:2020cgq}. In both cases, however, only the first few perturbative orders at weak coupling have been computed using the matrix model provided by the localization procedure \cite{Pestun:2007rz}. 
Our purpose here is, instead, to discuss these functions beyond perturbation theory in the large-$N$ limit and eventually find their strong-coupling behavior.
To this aim it is convenient to first recall a few properties of the quiver construction, 
most of them suggested by string theory.

\subsection{Fractional branes and twisted sectors}
\label{subsec:frac}

The quiver theory under consideration can be obtained with a $\mathbb{Z}_M$ 
orbifold projection starting from a parent $\cN=4$ SYM theory with gauge group SU($N$). This fact can be easily shown by taking Type II B string theory on a $\mathbb{C}^2/\mathbb{Z}_M$ orbifold singularity and a stack of $N$ regular D3-branes that engineer the $\cN=4$ SYM theory.
If we break this configuration into $M$ different stacks, each one containing $N$ fractional 
D3-branes located at the orbifold fixed-point, we obtain the quiver theory 
(see for instance \cite{Douglas:1996sw,Kachru:1998ys}). The fractional D3-branes carry an irreducible 
one-dimensional representation of $\mathbb{Z}_M$ and are associated to the nodes of the
quiver. Thus, they can be labeled by the same index $I$ introduced above.
In the field-theory limit, the massless open string starting and ending
on the $I$-th branes give rise to the adjoint vector multiplet of the
$I$-th node of the quiver, while the massless open strings stretching 
between the $I$-th branes and 
the $(I\pm1)$-th branes yield the bi-fundamental hypermultiplets.

From a geometrical point of view, the fractional D3-branes can be interpreted as D5-branes wrapped around the exceptional 2-cycles $e_i$ (with $i=1,\ldots,M-1$) of the $\mathbb{Z}_M$
orbifold singularity. These 2-cycles are associated to anti-self dual normalizable 2-forms 
$\omega^i$ in the sense that
\begin{equation}
\int_{e_i}\omega^j=\delta_i^{\,j}~,
\label{eomega}
\end{equation}
and are normalized in such a way that
\begin{equation}
\int_{\mathcal{M}}\omega^i\wedge\omega^j=-\big(C^{-1}\big)^{ij}~.
\label{Cij}
\end{equation}
Here $\mathcal{M}$ is the ALE space obtained by resolving the $\mathbb{C}^2/\mathbb{Z}_M$
orbifold singularity and $C$ is the Cartan matrix of the $\mathfrak{su}(M)$ algebra, namely
\begin{equation}
C=\begin{pmatrix}
2&-1&0&0&0&\ldots\\
-1&2&-1&0&0&\ldots\\
0&-1&2&-1&0&\ldots\\
\vdots&\vdots&\vdots&\vdots&\vdots&\ddots
\end{pmatrix}~.
\label{Cartanmatrix}
\end{equation}
Note that there are $M$ types of fractional branes but there are only $(M-1)$ 2-cycles $e_i$. 
In fact, the fractional branes corresponding to the trivial representation, {\it{i.e.}} those with 
$I=0$ in our conventions, are D5-branes wrapped around 
the 2-cycle $e_0=-\sum_i e_i$ (in presence
of an additional magnetic background flux on the world-volume). 

As extensively discussed in the literature (see for example \cite{Diaconescu:1999dt,Bertolini:2000dk,Billo:2001vg,Bertolini:2001qa,Ashok:2020jgb} and the review \cite{Bertolini:2001gq}), the fractional D3-branes admit a simple description from the closed string point of view in terms of ``boundary states''. These boundary states, denoted as
$|D3\rangle_{I}$, have a component $|U\rangle$, which corresponds to the untwisted sector of the closed string and is the same for all types of branes, and a component, which is a combination of the Ishibashi states $|T_\alpha\rangle$ corresponding 
to the twisted sectors of the closed string in the $\mathbb{Z}_M$
orbifold \cite{Dixon:1985jw,Dixon:1986jc} and is different for the different branes. 
In a very schematic notation, in which all inessential normalization factors are not exhibited, we have 
\begin{equation}
|D3\rangle_{I}=\frac{1}{\sqrt{M}}\bigg(|U\rangle+\sum_{\alpha=1}^{M-1}
\rho^{\alpha I}\,|T_\alpha\rangle\bigg)~.
\end{equation}
Inverting this formula, we obtain
\begin{subequations}
\begin{align}
|U\rangle&=\frac{1}{\sqrt{M}}\Big(|D3\rangle_{0}+|D3\rangle_{1}+\ldots
|D3\rangle_{M-1}\Big)~,\label{bsU}\\[1mm]
|T_\alpha\rangle&=\frac{1}{\sqrt{M}}\sum_{I=0}^{M-1} \rho^{-\alpha
I}\,|D3\rangle_{I}~.\label{bsTalpha}
\end{align}
\label{bsUT}%
\end{subequations}
Here we recognize exactly the same structure of the chiral primary operators (\ref{UT}),
and this fact explains why the operators $U_n$ have been called untwisted and the operators
$T_{\alpha,n}$ have been called twisted.

\subsection{Near-horizon description}
\label{subsec:nh}
{From} a geometrical point of view, the fractional D3-branes can be interpreted also as extended solitonic configurations of Type II B supergravity. In particular they emit the scalar fields corresponding to the wrapping of the 2-forms $B_{(2)}$ and $C_{(2)}$ of the Neveu-Schwarz/Neveu-Schwarz (NS/NS) and Ramond/Ramond (R/R) sectors, respectively, 
around the exceptional 2-cycles $e_i$ of the orbifold. 
This wrapping gives rise to the scalars\,%
\footnote{The factors of $\alpha^\prime$ have been inserted in order to make the
scalars dimensionless as the parent 2-forms.}
\begin{equation}
\hat{b}_i=\frac{1}{2\pi\alpha^\prime}\int_{e_i}B_{(2)}\quad\mbox{and}\quad
\hat{c}_i=\frac{1}{2\pi\alpha^\prime}\int_{e_i}C_{(2)}~,
\label{bIcI}
\end{equation}
where $\alpha^\prime$ is the square of the string length.
In addition, one can introduce also the following scalars\,%
\footnote{Here $B_{(2)}^\prime=B_{(2)}+2\pi\alpha^\prime\mathcal{F}$ where $\mathcal{F}$ is a constant background representing a unit magnetic flux.}
\begin{equation}
\hat{b}_0=\frac{1}{2\pi\alpha^\prime}\int_{e_0}B_{(2)}^\prime
\quad\mbox{and}\quad
\hat{c}_0=\frac{1}{2\pi\alpha^\prime}\int_{e_0}C_{(2)}
\label{relations}
\end{equation}
associated to the cycle $e_0=-\sum_ie_i$. Of course $b_0$ and $c_0$ are not independent fields, since
\begin{equation}
\hat{b}_0=1-\sum_{i=1}^{M-1}\hat{b}_i\quad\mbox{and}\quad
\hat{c}_0=-\sum_{i=1}^{M-1}\hat{c}_i~.
\end{equation}
Nevertheless, it is useful to consider them because, in complete analogy with (\ref{UT}) and 
(\ref{bsUT}), we can define the following untwisted and twisted combinations\,%
\footnote{With respect to (\ref{UT}) and (\ref{bsUT}), we have inserted overall factors 
of $1/2$ for later convenience.}
\begin{subequations}
\begin{align}
b&=\frac{1}{2\sqrt{M}}\big(\,\hat{b}_0+\hat{b}_1+\ldots \hat{b}_{M-1}\big)~,\qquad 
c=\frac{1}{2\sqrt{M}}\big(\,\hat{c}_0+\hat{c}_1+\ldots \hat{c}_{M-1}\big)~,\\[1mm]
b_\alpha&=
\frac{1}{2\sqrt{M}}\sum_{I=0}^{M-1} \rho^{-\alpha I}\,\hat{b}_{I}~,\qquad \qquad\quad\,
c_\alpha=\frac{1}{2\sqrt{M}}\sum_{I=0}^{M-1} \rho^{-\alpha I}\,\hat{c}_{I}~.
\label{bcIbcalpha}
\end{align}%
\end{subequations}
Note that, in view of (\ref{relations}), $b$ is constant and $c$ vanishes. Notice also that
$b_\alpha$ and $c_\alpha$ are complex fields, satisfying the following conjugation rules
\begin{equation}
b_\alpha^{\,*}=b_{M-\alpha}\quad\mbox{and}\quad
c_\alpha^{\,*}=c_{M-\alpha}~.
\end{equation}

To proceed, following \cite{Gukov:1998kk}, we consider the terms of the Type II B supergravity action in ten dimensions
that yield the linearized field equations for the 2-forms $B_{(2)}$ and $C_{(2)}$, namely
\begin{equation}
S_{10}=\frac{1}{2\kappa_{10}^2}\bigg[ \int d^{10}x\,\sqrt{G_{10}}~ \Big(\frac{1}{12}
\big(dB_{(2)}\,)^2+\frac{1}{12}\big(dC_{(2)}\,)^2\Big)-
\int \!4\,C_{(4)}\wedge dB_{(2)} \wedge dC_{(2)}\bigg]
\end{equation}
where
\begin{equation}
2\kappa_{10}^2=(2\pi)^7\,\alpha^{\prime\,4}\,g_s^2
\label{kappa10}
\end{equation}
is the gravitational constant in ten dimensions, $g_s$ is the string coupling,
$G_{10}$ is the determinant of the ten-dimensional metric and $C_{(4)}$ is the 4-form of the R/R
sector with a self-dual field strength $F_{(5)}$. We recall that in the presence of fractional D3-branes the dilaton and the R/R 0-form are both constant. Therefore, the only dilaton dependence in $S_{10}$ is through $g_s$ which is the exponential of the vacuum expectation value of the dilaton. Moreover, without any loss of generality, we can set the constant R/R 0-form to zero, so that the R/R 3-form field strength is just the exterior derivative of $C_{(2)}$. 

Wrapping the 2-forms on the exceptional cycles of the $\mathbb{Z}_M$ orbifold singularity, making use of (\ref{bIcI}), (\ref{eomega}) and (\ref{Cij}), and discarding a total derivative
term, we obtain from $S_{10}$ the following six-dimensional action
\begin{equation}
S_6=\frac{1}{2\kappa_{6}^2} \sum_{i,j=1}^{M-1}\bigg[
\int d^{6}x\,\sqrt{G_{6}}~ \Big(\frac{1}{2}\,
\partial \hat{b}_i \cdot\partial \hat{b}_j+\frac{1}{2}\,\partial \hat{c}_i \cdot\partial \hat{c}_j\Big)+
\int \!4\,F_{(5)}\wedge d\hat{b}_{i} \wedge \hat{c}_{j}
\bigg]\big(C^{-1}\big)^{ij}
\end{equation}
where
\begin{equation}
2\kappa_{6}^2=\frac{2\kappa_{10}^2}{(2\pi\alpha^\prime)^2}=(2\pi)^5\,\alpha^{\prime\,2}\,g_s^2
~.
\label{kappa6}
\end{equation}
Now we rewrite $\hat{b}_i$ and $\hat{c}_i$ in terms of the complex scalars $b_\alpha$ and $c_\alpha$ using the
inverse of (\ref{bcIbcalpha}) and, after some simple algebra, we find
\begin{equation}
S_6=\frac{1}{2} 
\sum_{\alpha=1}^{M-1}\frac{1}{2\kappa_{6}^2}\,\frac{1}{\sin^2\big(\frac{\pi\alpha}{M}\big)}\bigg[
\int \!d^{6}x\,\sqrt{G_{6}}~ \Big(
\partial b_{\alpha}^{\,*} \cdot\partial b_{\alpha}+
\partial c_{\alpha}^{\,*} \cdot\partial c_{\alpha}\Big)+
\int \!8\,F_{(5)}\wedge db_{\alpha}^{\,*} \wedge\, c_{\alpha}
\bigg]~.
\end{equation}
Up to this point the space-time geometry has not been specified (we only took into account the fact that the dilaton and the R/R 0-form are constant in the presence of fractional D3-branes). 
Now, instead,  following again \cite{Gukov:1998kk} we assume that the space where the 
scalars $b_\alpha$ and $c_\alpha$ propagate is of the form
\begin{equation}
\mathrm{AdS}_5\times S^1~.
\label{AdSS1}
\end{equation}
This represents the near-horizon geometry for the twisted fields. Indeed, as discussed in \cite{Kachru:1998ys}, the gravity dual of the quiver gauge theory realized
by the fractional D3-branes of $\mathbb{Z}_M$ is Type II B string theory on 
$\mathrm{AdS}_5 \times (S^5/\mathbb{Z}_M)$ in which the orbifold does not act on the AdS space but only on the 5-sphere. Then, it is easy to realize that the orbifold fixed point where the twisted
fields are located is precisely the six-dimensional space in (\ref{AdSS1}) with $S^1\subset S^5$.
Furthermore, the R/R 5-form $F_{(5)}$ is non-vanishing and is proportional to the volume form of
the AdS space.

We now insert this information in $S_6$ and perform a Kaluza-Klein compactification on 
$S^1$ by writing
\begin{equation}
b_\alpha=\frac{1}{\sqrt{2\pi}}\sum_{n\in \mathbb{Z}}b_{\alpha,n}\,\rme^{\ii n\theta} 
\quad\mbox{and}\quad
c_\alpha=\frac{1}{\sqrt{2\pi}}\sum_{n\in \mathbb{Z}}c_{\alpha,n}\,\rme^{\ii n\theta} 
\end{equation}
where $\theta$ is the coordinate of the circle and the Fourier modes are functions only on $\mathrm{AdS}_5$ which satisfy the following complex conjugation rules
\begin{equation}
b_{\alpha,n}^{\,*}=b_{M-\alpha,-n}\quad\mbox{and}\quad
c_{\alpha,n}^{\,*}=c_{M-\alpha,-n}~.
\end{equation}
In this way, the action $S_6$ reduces to
\begin{equation}
S_{\mathrm{AdS}_5}=\frac{1}{2}\sum_{\alpha=1}^{M-1}
\sum_{n\in\mathbb{Z}}\,
\int_{\mathrm{AdS}_5} \!\!d^5x\, 
\sqrt{G_{\mathrm{AdS}_5}}~\mathcal{L}_{\alpha,n}
\end{equation}
where the Lagrangian $\mathcal{L}_{\alpha,n}$ is
\begin{equation}
\mathcal{L}_{\alpha,n}=\frac{1}{2\kappa_{6}^2}\,\frac{1}{\sin^2\big(\frac{\pi\alpha}{M}\big)}\,
\Big(b_{\alpha,n}^{\,*}\,,\,c_{\alpha,n}^{\,*}\Big)\cdot
\begin{pmatrix}
-\Delta+n^2&-4\ii n\\
4\ii n&-\Delta+n^2
\end{pmatrix}\cdot\begin{pmatrix}
b_{\alpha,n}\\
c_{\alpha,n}
\end{pmatrix}
\end{equation}
with $\Delta$ being the Laplace operator in ${\mathrm{AdS}_5}$.
It is convenient to diagonalize the quadratic form and introduce the following combinations
\begin{subequations}
\begin{align}
\gamma_{\alpha,n}&=c_{\alpha,n}+\ii\,b_{\alpha,n}~.\\
\eta_{\alpha,n}&=c_{\alpha,n}-\ii\,b_{\alpha,n}~,
\end{align}
\label{gammaeta}%
\end{subequations}
which are eigenvectors of the mass matrix corresponding to the following eigenvalues
\begin{subequations}
\begin{align}
m_{\gamma_{\alpha,n}}^2&=n(n+4)~,\label{mgamma}\\
m_{\eta_{\alpha,n}}^2&=n(n-4)~.\label{meta}
\end{align}
\label{mass}%
\end{subequations}
As discussed in \cite{Gukov:1998kk}, this mass spectrum perfectly accounts for the scalar
operators of the quiver gauge theory. In particular, applying the AdS/CFT dictionary \cite{Gubser:1998bc,Witten:1998qj}, we see from
(\ref{meta}) that the modes $\eta_{\alpha,n}$ are dual to conformal operators of dimension 
$n$ and these are precisely the twisted operators $T_{\alpha,n}(\vec{x})$ defined in (\ref{Tnalpha})
for $n\geq 2$. Therefore, in order to obtain information on the correlation functions of these operators at strong-coupling within the AdS/CFT correspondence, one can use the following action 
on the AdS side
\begin{equation}
S[\eta]=\frac{1}{2}\sum_{\alpha=1}^{M-1}
\sum_{n=2}^{\infty}\,
\int_{\mathrm{AdS}_5} \!\!d^5x\, 
\sqrt{G_{\mathrm{AdS}_5}}\,\bigg[\frac{1}{2\kappa_{6}^2}\,\frac{1}{\sin^2\big(\frac{\pi\alpha}{M}\big)}\,\Big(\partial\eta_{\alpha,n}^{\,*}\cdot\partial
\eta_{\alpha,n}+n(n-4)\,\eta_{\alpha,n}^{\,*}\,\eta_{\alpha,n}\Big)\bigg]~,
\label{Seta}
\end{equation}
supplemented by a term describing the coupling between $\eta_{\alpha,n}$ and
$T_{\alpha,n}$ on the boundary of $\mathrm{AdS}_5$ which is proportional to
\begin{equation}
\sum_{\alpha=1}^{M-1}
\sum_{n=2}^{\infty}\,
\int_{\partial(\mathrm{AdS}_5)} \!\!d^4x\,\Big[T_{\alpha,n}(\vec{x})\,\eta_{\alpha,n}(\vec{x})
+\,\mathrm{c.c.}\,\Big]~.
\end{equation}

We conclude by observing that the scalar modes $\gamma_{\alpha,n}$ are dual to conformal operators of dimension $(n+4)$ as one can see from (\ref{mgamma}). Such operators can be obtained
starting from
\begin{equation}
\tr \Big[\big(F_{I}^2+\ii \,F_{I}\,\widetilde{F}_{I}\big)\Phi_{I}(\vec{x})^n\Big]
\end{equation}
for $n\geq 0$, where $F_{I}$ and $\widetilde{F}_{I}$ are, respectively, 
the gauge field strength and its dual \cite{Gukov:1998kk}, 
and summing them over all nodes of the quiver with 
weights $\rho^{-\alpha I}$ in analogy with what we did in (\ref{Tnalpha}) with the
operators $O_{n}^{(I)}(\vec{x})$. When $n=0$, the field $\gamma_{\alpha,0}$
is dual to the (complexified) gauge coupling constants of the quiver model, 
and this fact is the near-horizon counterpart of what was found in \cite{Bertolini:2000dk,Polchinski:2000mx,Billo:2001vg,Bertolini:2001qa} when the geometry produced by the 
fractional D3-branes was analyzed in the probe approximation.

\section{Perturbative analysis from Feynman diagrams}
\label{secn:diagrams}
In this section we give a derivation of the first perturbative contributions to the 2-point
functions (\ref{UUTT}) using Feynman diagrams. In order to be brief, we present only the essential ingredients of the calculation and refer to \cite{Galvagno:2020cgq,Galvagno:2021bbj} for details 
and further elements.

\subsection{Feynman rules}
To write the Feynman rules it is convenient to decompose the field 
content of the $\cN=2$ quiver theory in terms of $\cN=1$ superfields as follows:
\begin{equation}
\begin{aligned}
&\cN=2~\mbox{vector superfield in the $I$-th node}~=\big(V_I,\Phi_I)~,\\
&\cN=2~\mbox{hypermultiplet connecting the $I$- and $(I+1)$-th nodes}~=
 \big(Q_I, \widetilde{Q}_I \big)~,
\end{aligned}
\label{N2fields}
\end{equation}
where $V_I$ is a $\cN=1$ vector superfield, while $\Phi_I$, $Q_I$ and $\widetilde{Q}_I$ 
are $\cN=1$ chiral superfields\,%
\footnote{With an abuse of notation we denote by $\Phi_I$ both the chiral superfield and its
lowest component, namely the complex scalar used in Section~\ref{sec:model}.}.
The multiplets $V_I$ and $\Phi_I$ transform in the adjoint representation 
of $\mathrm{SU}(N)_I$, while $Q_I$ and $\widetilde{Q}_I$ are in the bi-fundamental 
of $\mathrm{SU}(N)_{I} \times \mathrm{SU}(N)_{I+1}$. Thus, they can be written as\,%
\footnote{In the following formulas the indices $a,b$ are adjoint indices, the (upper) lower indices $A,B$ are (anti-)bi-fundamental indices, while 
the indices $u,v,\hat u,\hat v$ are fundamental indices.}
\begin{equation}
V_I = V_I^a (T_a)^u_{~v}~,\qquad
\Phi_I = \Phi_I^a (T_a)^u_{~v}~,\qquad  
Q_I=Q_I^A(B_A)^u_{~\hat{v}}~,\qquad
\widetilde{Q}_I = \widetilde{Q}_{I,A}(B^A)^{\hat{u}}_{~v}
\end{equation}
where $T_a$ are the generators of $\mathfrak{su}(N)$, while $B_A$ and $B^A$ 
are such that
$(B_A)^u_{~\hat{v}}(B^A)^{\hat{u}}_{~v} = \delta^u_v \,\delta^{\hat u}_{\hat v}$.

{From} the action of the generic quiver theory, which can be found for example in
Section~2 of \cite{Galvagno:2020cgq}, one can derive the Feynman rules. Those that will
be useful for us are:
\begin{subequations}
\begin{align}
V_I\,\Phi^\dagger_I\,\Phi_I-\mathrm{vertex}~:\parbox[c]{.2\textwidth}{\includegraphics[width = .2\textwidth]{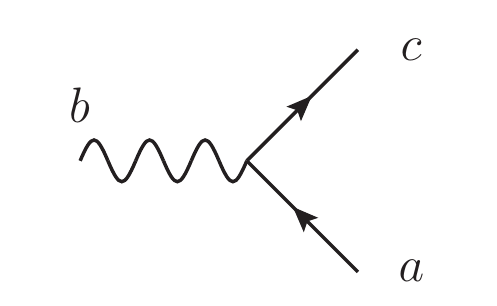}} &= 4\, g\, \tr \big(T_a \big[T_b,T_c\big]\Big)~,  
\\[1mm]
\Phi_I\,Q_I\,\widetilde{Q}_I ~~\mathrm{or}~~\Phi_{I+1}\,Q_I\widetilde{Q}_I
-\mathrm{vertex}~:\parbox[c]{.2\textwidth}{\includegraphics[width = .2\textwidth]{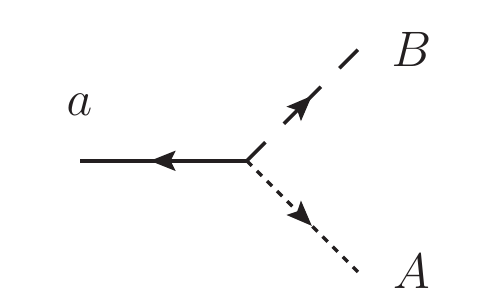}} &= \ii\,\sqrt{2}\,g\, \tr \big(T^a B^A B_B\big)\,\bar{\theta}^2~, 
\\[1mm]
{\Phi}_I^\dagger\,{Q}_I^\dagger\,{\widetilde{Q}}_I^\dagger
~~\mathrm{or}~~{\Phi}_{I+1}^\dagger\,{Q}_I^\dagger\,{\widetilde{Q}}_I^\dagger
-\mathrm{vertex}~:\parbox[c]{.2\textwidth}{\includegraphics[width = .2\textwidth]{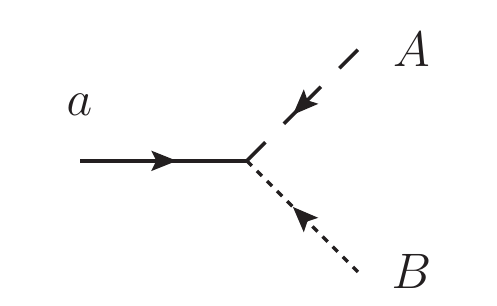}} &= -\ii\,\sqrt{2}\,g\, \tr \big(T^a B^A B_B\big)\,\theta^2
\end{align}
\label{vertices}%
\end{subequations}
where $\theta$ and $\overbar{\theta}$ are the super-space coordinates.
With these Feynman rules it is possible to compute the first perturbative contributions to the
correlation functions of the gauge invariant operators (\ref{OIn}) of the quiver theory, as we are going to discuss.

\subsection{Twisted and untwisted correlators at the perturbative level}

\begin{figure}[ht]
\begin{center}
\includegraphics[scale=0.7]{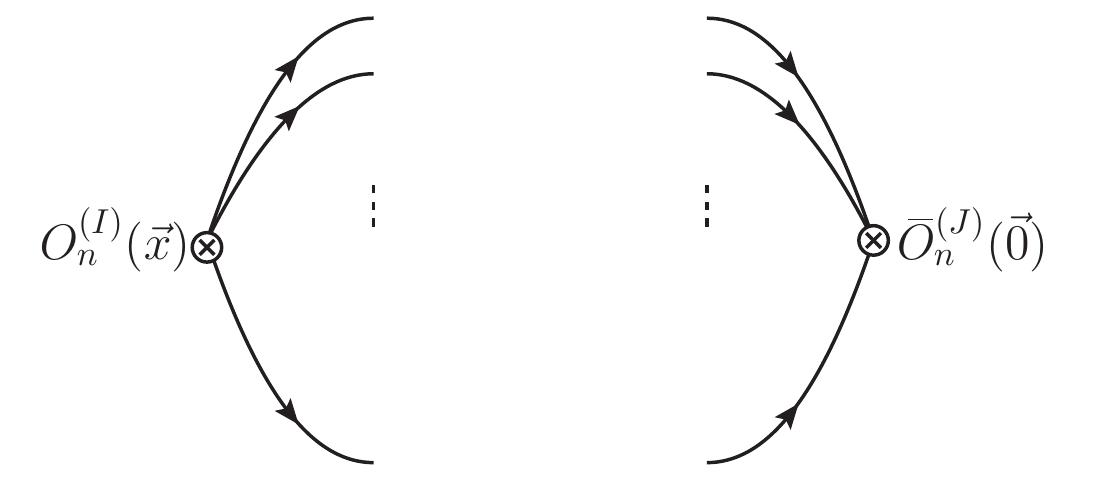}
\end{center}
\caption{The diagram representing the insertion of a chiral operator $O_{n}^{(I)}$ from the $I$ -th node in $\vec{x}$ and of the anti-chiral operator $\overbar{O}_{n}^{(J)}$ from the $J$ -th node in the origin. The number of propagators connecting the insertion points is $n$.}
\label{Fig:tree}
\end{figure}

The starting point for the evaluation of the chiral/anti-chiral 2-point functions is the 
vacuum expectation value 
$\big\langle O_n^{(I)}(\vec{x})\, \overbar{O}_n^{(J)}(\vec{0})\big\rangle$, represented
by the diagram in Fig.~\ref{Fig:tree}.

In the $\cN=4$ SYM theory the full result is obtained by closing the above diagram with $n$ tree-level scalar propagators, since in this case all loop corrections vanish \cite{Lee:1998bxa}. We denote 
the $\cN=4$ correlators as $\big\langle
O_n(\vec{x})\, \overbar{O}_n(\vec{0})\big\rangle_{0}$.
In $\cN=2$ theories, like the quiver model we are considering, there are instead perturbative contributions at all loops. However, in the large-$N$ limit the number of terms that remain at leading order drastically reduces and one is left with a few building blocks that contribute in a simple and controlled way. 

Using the results of Section 5 of \cite{Galvagno:2020cgq}, one can show that at large $N$
the non-trivial contributions to the 2-point function $\big\langle
O_n^{(I)}(\vec{x})\, \overbar{O}_n^{(J)}(\vec{0})\big\rangle$ arise only from diagrams that 
contain the following sub-diagrams, built using the vertices (\ref{vertices}):
\begin{align}
\label{Davydichev1a}
\parbox[c]{.15\textwidth}{\includegraphics[width = .15\textwidth]{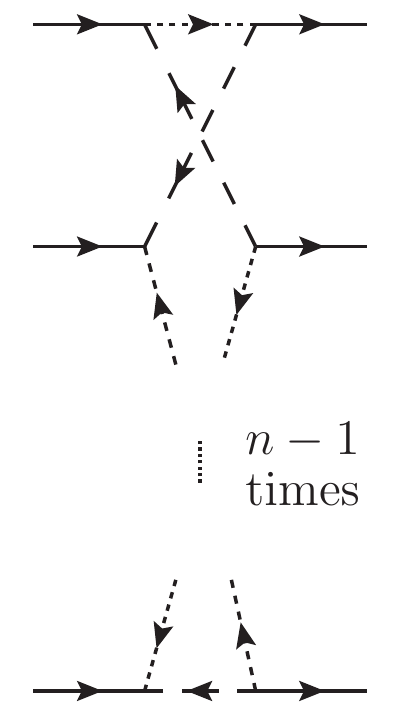}} \qquad
=&~~ \Big(\frac{-g^2}{16\pi^2}\Big)^{n} \,\binom{2n}{n} \,\frac{\zeta(2n-1)}{n} \,\times \,
\Big(\frac{1}{4\pi^2 \vec{x}^{\,2}}\Big)^n ~,
\end{align}
and
\begin{align}
\label{Davydichev1b}
\parbox[c]{.16\textwidth}{\includegraphics[width = .16\textwidth]{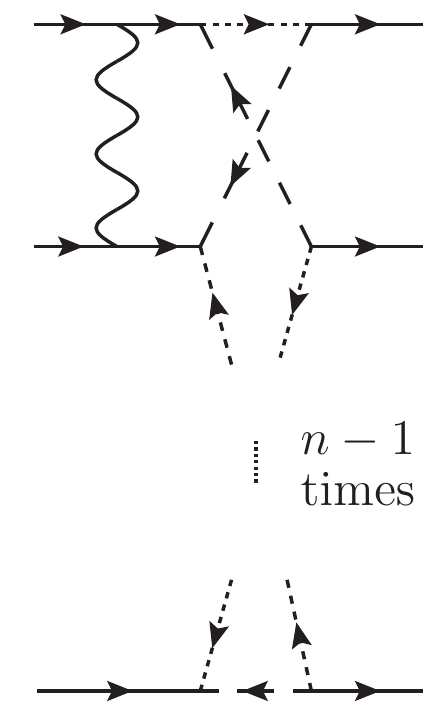}} \qquad~~
=&~~ \Big(\frac{-g^2}{16\pi^2}\Big)^{n+1} \,n\,\binom{2n+2}{n+1} \,\frac{\zeta(2n+1)}{n} 
\,\times \, \Big(\frac{1}{4\pi^2 \vec{x}^{\,2}}\Big)^n ~.
\end{align}

By inserting these structures inside the diagram of Fig.~\ref{Fig:tree} in all possible ways, one obtains the perturbative contributions to the 2-point functions $\big\langle
O_n^{(I)}(\vec{x})\, \overbar{O}_n^{(J)}(\vec{0})\big\rangle$. 
In particular, when $I=J$ the first corrections arise from a single insertion of
(\ref{Davydichev1a}) and (\ref{Davydichev1b}). Normalizing with respect to the correlators of the
$\cN=4$ SU($N$) SYM theory, at large $N$ we find
\begin{align}
\label{sameZq}
\frac{\big\langle O_{n}^{(I)}(\vec{x})\, \overbar{O}_{n}^{(I)}(\vec{0})\big\rangle\phantom{\Big{|}}}{~\big\langle 
O_{n}(\vec{x})\, \overbar{O}_{n}(\vec{0})\big\rangle_{0}\phantom{\Big{|}}} =
1-\Big(\frac{\lambda}{8\pi^2}\Big)^{n}\binom{2n}{n}\,\frac{\zeta(2n-1)}{2^{n-1}}
+\Big(\frac{\lambda}{8\pi^2}\Big)^{n+1}\binom{2n+2}{n+1}\,\frac{n\,\zeta(2n+1)}{2^{n-1}}+\ldots
\end{align}
where we have introduced the 't Hooft coupling
\begin{equation}
\lambda=N\,g^2
\label{lambda}
\end{equation}
which is kept fixed when $N\to\infty$.

When $J=I\pm1$, again a single insertion of the building blocks
(\ref{Davydichev1a}) and (\ref{Davydichev1b}) yields the first perturbative terms, which
read\,%
\footnote{Notice that an extra factor of 2 must be included in the $M=2$ theory, in order
to properly take into account the fact that the two nodes are nearest neighbors of each other.
\label{footnotefactor2}} 
\begin{align}
\label{neighborsZq}
\frac{\big\langle O_{n}^{(I)}(\vec{x})\, \overbar{O}_{n}^{(I\pm1)}(\vec{0})\big\rangle\phantom{\Big{|}}}{~\big\langle 
O_{n}(\vec{x})\, \overbar{O}_{n}(\vec{0})\big\rangle_{0}\phantom{\Big{|}}} 
=\Big(\frac{\lambda}{8\pi^2}\Big)^{n}\binom{2n}{n}\,\frac{\zeta(2n-1)}{2^{n}}
-\Big(\frac{\lambda}{8\pi^2}\Big)^{n+1}\binom{2n+2}{n+1}
\,\frac{n\,\zeta(2n+1)}{2^{n}}+\dots~.
\end{align}

\begin{figure}[ht]
\begin{center}
\includegraphics[scale=0.45]{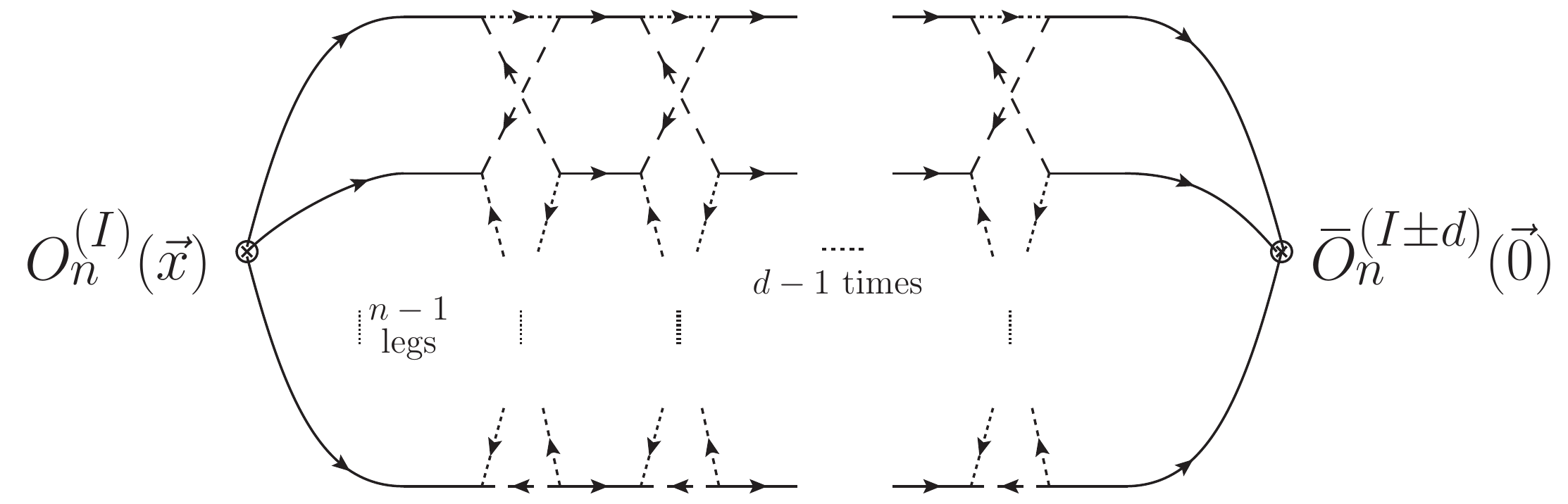}
\end{center}
\caption{The diagram describing the leading contribution to the 2-point function $\big\langle O_{n}^{(I)}(\vec{x})\, \overbar{O}_{n}^{(I\pm d)}(\vec{0})\big\rangle$. This contribution is proportional to $\lambda^{nd}\,\zeta(2n-1)^d$.}
\label{Fig:QdLO}
\end{figure}

\begin{figure}[ht]
\begin{center}
\includegraphics[scale=0.45]{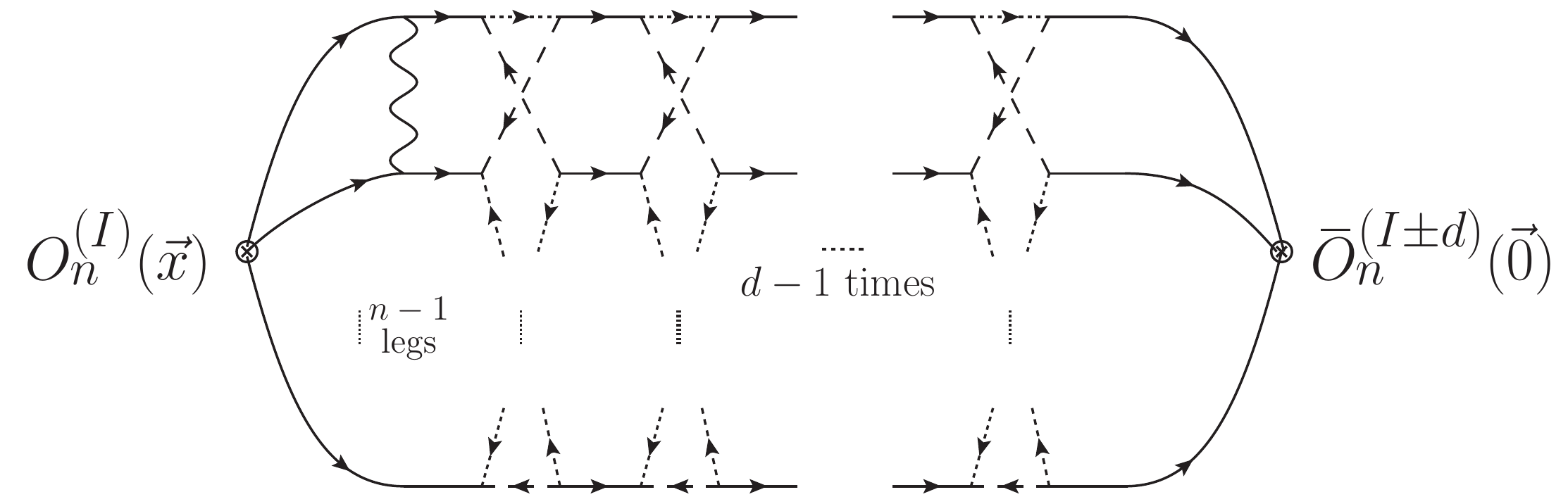}
\end{center}
\caption{The diagram describing the next-to-leading contribution to $\big\langle O_{n}^{(I)}(\vec{x})\, \overbar{O}_{n}^{(I\pm d)}(\vec{0})\big\rangle$. This contribution is proportional to 
$\lambda^{nd+1}\,\zeta(2n-1)^{d-1}\,\zeta(2n+1)$.}
\label{Fig:QdNLO}
\end{figure}

When $J=I\pm d$ with $d>1$, one needs to consider multiple insertions of the sub-diagrams
(\ref{Davydichev1a}) and (\ref{Davydichev1b}) in order to connect the operators that are distant
$d$ nodes from each other. Using again the results of \cite{Galvagno:2020cgq}, one can prove that
the leading contribution to the 2-point functions corresponds to the diagram of Fig.~\ref{Fig:QdLO}
which contains $d$ insertions of the sub-diagram (\ref{Davydichev1a}) and thus is proportional to
$\lambda^{nd}\,\zeta(2n-1)^d$.
The next-to-leading contribution arises instead from one insertion of the sub-diagram (\ref{Davydichev1b}) and $(d-1)$ insertions of the sub-diagram (\ref{Davydichev1a}), as represented in Fig.~\ref{Fig:QdNLO}.
This contribution is proportional to $\lambda^{nd+1}\,\zeta(2n-1)^{d-1}\,\zeta(2n+1)$
and thus has a power of $\lambda$ more than the previous one. These results show that
in order to find the perturbative terms in the 2-point correlators that are leading in $\lambda$ and linear in the Riemann $\zeta$-values, it is enough to consider 
the contributions from operators inserted in the same node or in neighboring nodes, 
given respectively in (\ref{sameZq}) and (\ref{neighborsZq}).

\subsubsection{Examples}
Let us now consider some explicit examples.
\paragraph{The $M=2$ theory:}
For $M=2$, the untwisted and twisted operators are given by (\ref{UT}) with $\rho=-1$. Using these
expressions and the fact that the correlators (\ref{sameZq}) and (\ref{neighborsZq}) do not depend
on the value of $I$, it is straightforward to show that
\begin{equation}
\begin{aligned}
\big\langle U_n^{\phantom{\dagger}}(\vec{x})\,{U}_n^{\,\dagger}(\vec{0})\big\rangle&=
\big\langle O_n^{(0)}(\vec{x})\,\overbar{O}_n^{(0)}(\vec{0})\big\rangle+
\big\langle O_n^{(0)}(\vec{x})\,\overbar{O}_n^{(1)}(\vec{0})\big\rangle~,\\[1mm]
\big\langle T_{1,n}^{\phantom{\dagger}}(\vec{x})\,{T}_{1,n}^{\,\dagger}(\vec{0})\big\rangle&=
\big\langle O_n^{(0)}(\vec{x})\,\overbar{O}_n^{(0)}(\vec{0})\big\rangle-
\big\langle O_n^{(0)}(\vec{x})\,\overbar{O}_n^{(1)}(\vec{0})\big\rangle~.
\end{aligned}
\end{equation}
Then, taking into account the observation in the footnote~\ref{footnotefactor2}, 
in the large-$N$ limit we have
\begin{align}
\frac{\big\langle U_n^{\phantom{\dagger}}(\vec{x})\,{U}_n^{\,\dagger}(\vec{0})\big\rangle\phantom{\Big{|}}}{~\big\langle 
O_{n}(\vec{x})\, \overbar{O}_{n}(\vec{0})\big\rangle_{0}\phantom{\Big{|}}} &=1+\ldots~,
\label{UnUn}
\end{align}
where the ellipses stand for possible higher-order terms, and
\begin{align}
\frac{\big\langle T_{1,n}^{\phantom{\dagger}}(\vec{x})\,{T}_{1,n}^{\,\dagger}(\vec{0})\big\rangle\phantom{\Big{|}}}{~\big\langle 
O_{n}(\vec{x})\, \overbar{O}_{n}(\vec{0})\big\rangle_{0}\phantom{\Big{|}}} &=1
-\Big(\frac{\lambda}{8\pi^2}\Big)^{n}\binom{2n}{n}\,\frac{\zeta(2n-1)}{2^{n-2}}
+\Big(\frac{\lambda}{8\pi^2}\Big)^{n+1}\binom{2n+2}{n+1}\,\frac{n\,\zeta(2n+1)}{2^{n-2}}+\ldots~.
\label{T1nT1n}
\end{align}
It is worth pointing out that in the untwisted correlator there is an exact 
cancellation of the first perturbative contributions (which actually persists also at higher orders). 
This implies that in the planar limit these untwisted correlators exactly match those of the $\cN=4$ SYM theory. On the contrary, the twisted correlators differ from the $\cN=4$ ones already at the planar level, with the first correction being proportional to $\lambda^n\,\zeta(2n-1)$.

\paragraph{The $M=3$ theory:}
For $M=3$, the untwisted and twisted operators are given by (\ref{UT}) with 
$\rho=\rme^{\frac{2\pi\ii}{3}}$. In this case we have two twisted sectors that are conjugate to each other, and the corresponding operators are $T_{1,n}$ and $T_{2,n}$. 
Proceeding as before, we easily find
\begin{equation}
\begin{aligned}
\big\langle U_n^{\phantom{\dagger}}(\vec{x})\,{U}_n^{\,\dagger}(\vec{0})\big\rangle&=
\big\langle O_n^{(0)}(\vec{x})\,\overbar{O}_n^{(0)}(\vec{0})\big\rangle+2\,
\big\langle O_n^{(0)}(\vec{x})\,\overbar{O}_n^{(1)}(\vec{0})\big\rangle~,\\[2mm]
\big\langle T_{\alpha,n}^{\phantom{\dagger}}(\vec{x})\,{T}_{\alpha,n}^{\,\dagger}(\vec{0})\big\rangle
&=\big\langle O_n^{(0)}(\vec{x})\,\overbar{O}_n^{(0)}(\vec{0})\big\rangle-
\big\langle O_n^{(0)}(\vec{x})\,\overbar{O}_n^{(1)}(\vec{0})\big\rangle
\qquad\mbox{for}~\alpha=1,2~.
\end{aligned}
\end{equation}
Using (\ref{sameZq}) and (\ref{neighborsZq}) in the first line, we see that the first perturbative contributions cancel each other, so that the untwisted correlators are again as in (\ref{UnUn}). 
For the twisted correlators, we obtain
\begin{equation}
\frac{\big\langle T_{\alpha,n}^{\phantom{\dagger}}(\vec{x})\,{T}_{\alpha,n}^{\,\dagger}(\vec{0})\big\rangle\phantom{\Big{|}}}{~\big\langle 
O_{n}(\vec{x})\, \overbar{O}_{n}(\vec{0})\big\rangle_{0}\phantom{\Big{|}}} =1
-\frac{3}{4}\Big(\frac{\lambda}{8\pi^2}\Big)^{n}\binom{2n}{n}\,\frac{\zeta(2n-1)}{2^{n-2}}
+\frac{3}{4}\Big(\frac{\lambda}{8\pi^2}\Big)^{n+1}\binom{2n+2}{n+1}\,\frac{n\,\zeta(2n+1)}{2^{n-2}}+\ldots
\label{T1nT2n}
\end{equation}
for $\alpha=1,2$.
This is the same structure found in (\ref{T1nT1n}), with just a different numerical coefficient
in front of the perturbative terms: 3/4 instead of 1.

\paragraph{The $M=4$ theory:} For $M=4$, the untwisted and twisted operators are given by (\ref{UT}) with $\rho=\ii$. In this case we have two twisted sectors that are conjugate to each other, whose operators are $T_{1,n}$ and $T_{3,n}$, and one self-conjugate twisted sector with
operators $T_{2,n}$. Following the same steps described above, we find
\begin{equation}
\begin{aligned}
\big\langle U_n^{\phantom{\dagger}}(\vec{x})\,{U}_n^{\,\dagger}(\vec{0})\big\rangle&=
\big\langle O_n^{(0)}(\vec{x})\,\overbar{O}_n^{(0)}(\vec{0})\big\rangle+2\,
\big\langle O_n^{(0)}(\vec{x})\,\overbar{O}_n^{(1)}(\vec{0})\big\rangle+\ldots~,\\[2mm]
\big\langle T_{\alpha,n}^{\phantom{\dagger}}(\vec{x})\,{T}_{\alpha,n}^{\,\dagger}(\vec{0})\big\rangle&=
\big\langle O_n^{(0)}(\vec{x})\,\overbar{O}_n^{(0)}(\vec{0})\big\rangle+\ldots\qquad\mbox{for}
~\alpha=1,3~,\\[2mm]
\big\langle T_{2,n}^{\phantom{\dagger}}(\vec{x})\,{T}_{2,n}^{\,\dagger}(\vec{0})\big\rangle&=
\big\langle O_n^{(0)}(\vec{x})\,\overbar{O}_n^{(0)}(\vec{0})\big\rangle-2
\big\langle O_n^{(0)}(\vec{x})\,\overbar{O}_n^{(1)}(\vec{0})
\big\rangle+\ldots
\end{aligned}
\label{UTZ4}
\end{equation}
where the ellipses stand for contributions from operators in non-nearest neighbor nodes that yield
higher-order corrections with multiple Riemann $\zeta$-values. 

{From} the first line of (\ref{UTZ4}) we see that the untwisted correlators
are as in (\ref{UnUn}), while the second and third lines tell us that the
twisted 2-point functions are given by
\begin{equation}
\frac{\big\langle T_{\alpha,n}^{\phantom{\dagger}}(\vec{x})\,{T}_{\alpha,n}^{\,\dagger}(\vec{0})\big\rangle\phantom{\Big{|}}}{~\big\langle 
O_{n}(\vec{x})\, \overbar{O}_{n}(\vec{0})\big\rangle_{0}\phantom{\Big{|}}} =1
-\frac{1}{2}\Big(\frac{\lambda}{8\pi^2}\Big)^{n}\binom{2n}{n}\,\frac{\zeta(2n-1)}{2^{n-2}}
+\frac{1}{2}\Big(\frac{\lambda}{8\pi^2}\Big)^{n+1}\binom{2n+2}{n+1}\,\frac{n\,\zeta(2n+1)}{2^{n-2}}+\ldots
\label{TnZ4a}
\end{equation}
for $\alpha=1,3$, and
\begin{equation}
\frac{\big\langle T_{2,n}^{\phantom{\dagger}}(\vec{x})\,{T}_{2,n}^{\,\dagger}(\vec{0})\big\rangle\phantom{\Big{|}}}{~\big\langle 
O_{n}(\vec{x})\, \overbar{O}_{n}(\vec{0})\big\rangle_{0}\phantom{\Big{|}}} =1
-\Big(\frac{\lambda}{8\pi^2}\Big)^{n}\binom{2n}{n}\,\frac{\zeta(2n-1)}{2^{n-2}}
+\Big(\frac{\lambda}{8\pi^2}\Big)^{n+1}\binom{2n+2}{n+1}\,\frac{n\,\zeta(2n+1)}{2^{n-2}}+\ldots~.
\label{TnZ4b}
\end{equation}
Again we have the same structure of (\ref{T1nT1n}) and (\ref{T1nT2n}), but with different
overall coefficients in the perturbative terms, namely 1/2 and 1 in this case.

It is clear from these explicit examples that even if the operators $O_n^{(I)}$ seem to be the natural objects to be studied, it is actually far more convenient to consider the combinations $U_n$ and $T_{\alpha,n}$ defined in
(\ref{UT}) and suggested by the string construction reviewed in the previous section. Indeed, in the correlators of two untwisted operators there are exact cancellations of the perturbative contributions, as in the $\cN=4$ SYM theory, while the correlators of two twisted operators always
exhibit the same structure, in which the difference between the quiver theories is encoded in simple numerical coefficients. Of course to make real progress in this analysis, it is necessary to compute higher-order perturbative contributions and check whether these properties persist. However, the diagrammatic approach we have discussed, although simple in principle, rapidly becomes unpractical when the number of loops increases, and new techniques have to be used.

\section{The matrix model}
\label{sec:matrix}
A very efficient way to compute the 2-point functions of chiral/anti-chiral operators in $\cN=2$ gauge theories, which can be easily extended to very high orders in perturbation theory, 
is the one based on the localization technique \cite{Pestun:2016jze} that maps 
the field theory to an interacting matrix model \cite{Pestun:2007rz} reducing the calculation of correlation functions to finite-dimensional matrix integrals. These integrals can be easily computed 
in the so-called full Lie algebra approach, in which one integrates over all matrix elements and uses recursion relations \cite{Billo:2017glv} to obtain explicit expressions both at finite $N$ and at large $N$. The whole procedure can be nicely implemented in an algorithmic way, as shown in \cite{Beccaria:2020hgy,Galvagno:2020cgq,Beccaria:2021hvt}, so that very long perturbative expansions can be generated with little effort.

In the following we provide the basic ingredients that are necessary to perform these calculations.

\subsection{Explicit form of the quiver matrix model}
\label{subsec:explmatmodel}
When the quiver theory is placed on a four-sphere of unit radius, its partition function localizes \cite{Pestun:2007rz} and can be written as an integral of a multi-matrix model defined in terms 
of $N\times N$ traceless matrices $a_{I}$ (with $I=0,\ldots, M-1$) as follows
\begin{equation}
\mathcal{Z}=\int \prod_{I=0}^{M-1} \Big( da_I\,\,\rme^{-\tr a_I^2}\Big)\,
\big|Z_{\mathrm{1-loop}}\,Z_{\mathrm{inst}}\big|^2~.
\label{ZS4}
\end{equation}
Here the factor $Z_{\mathrm{1-loop}}$ accounts for the 1-loop fluctuations around the localization
fixed points, while the $Z_{\mathrm{inst}}$ is the non-perturbative instanton contribution. Since we
are primarily interested in the large-$N$ 't Hooft limit in which instantons are suppressed, we put $Z_{\mathrm{inst}}=1$. 

In the full Lie algebra approach we integrate over all matrix elements. More explicitly, writing
\begin{equation}
a_I=a_I^b\,T_b
\end{equation}
where $T_b$ are the $\mathfrak{su}(N)$ generators satisfying
$\tr T_b\,T_c=\frac{1}{2}\delta_{bc}$, the integration measure in (\ref{ZS4}) is
\begin{equation}
da_I=\prod_{b=1}^{N^2-1}\,\frac{da_I^b}{\sqrt{2\pi}}
\label{measure}
\end{equation}
where the normalization has been chosen in such a way that the Gaussian integration for each $I$ yields 1.

As shown in \cite{Galvagno:2020cgq}, the 1-loop factor can be written as
\begin{equation}
\big|Z_{\mathrm{1-loop}}\big|^2=\rme^{-S_{\mathrm{int}}}
\label{Z1loop}
\end{equation}
with
\begin{align}
	\label{sinta}
		S_{\mathrm{int}} & = \sum_{I=0}^{M-1} \!\bigg[\sum_{m=2}^\infty \sum_{k=2}^{2m}
		(-1)^{m+k} \Big(\frac{g^2}{8\pi^2}\Big)^{\!m} \,
		\binom{2m}{k}\,\frac{\zeta(2m-1)}{2m} 
		\big(\tr a_{I}^{2m-k} - \tr a_{I+1}^{2m-k}\,\big) \big(\tr a_{I}^{k} - \tr a_{I+1}^{k}\big)\bigg]
		~.
\end{align}
Then, the partition function (\ref{ZS4}) becomes
\begin{equation}
\mathcal{Z}=\int \prod_{I=0}^{M-1} \Big( da_I\,\,\rme^{-\tr a_I^2}\Big)\,\rme^{-S_{\mathrm{int}}}
\label{ZS41}
\end{equation}
and $S_{\mathrm{int}}$ can be interpreted as an interaction action.

\subsubsection*{Untwisted and twisted operators}
\label{subsubsec:twistedops}
In the multi-matrix model introduced above, a natural basis of ``gauge invariant'' operators is provided by the traces of powers of the matrices in each node, namely $\tr a_{I}^n$. However, the analysis of the previous
two sections suggests that these operators should be reorganized into untwisted and twisted combinations, analogously to what we did in (\ref{UT}) for the field theory operators. 
Thus, for any $n\geq 2$ we introduce the untwisted operators
\begin{align}
	\label{untA}
		A_{n} = \frac{1}{\sqrt{M}} \Big(\tr a_{0}^n+\tr a_{1}^n+\ldots+\tr a_{M-1}^n\Big)~,
\end{align}
and the twisted ones
\begin{align}
	\label{twal}
		A_{\alpha, n} = \frac{1}{\sqrt{M}} \sum_{I=0}^{M-1}\, \rho^{-\alpha I}\, \tr a_{I}^n
\end{align}
where $\alpha=1,\ldots,M-1$. The untwisted operators $A_n$ are real, while the twisted ones satisfy
\begin{align}
	\label{realtau}
		A_{\alpha,n}^\dagger = A_{M-\alpha,n}~.
\end{align}
Note that, for $M$ even, the operators $A_{M/2,n}$ are real. For example, in the $\mathbb{Z}_3$ 
quiver theory we have
\begin{equation}
\begin{aligned}
	\label{z3tw}
		A_{n} & = \frac{1}{\sqrt{3}} \Big(\tr a_{0}^n + \tr a_{1}^n + \tr a_{2}^n\Big)~,\\
		A_{1, n}  & = \frac{1}{\sqrt{3}} \Big(\tr a_{0}^n + \rho^2 \tr a_{1}^n + \rho \tr a_{2}^n\Big)~,\\
		A_{2, n}  & = \frac{1}{\sqrt{3}} \Big(\tr a_{0}^n + \rho \tr a_{1}^n + \rho^2 \tr a_{2}^n\Big)~,
\end{aligned}
\end{equation}
where $\rho=\rme^{\frac{2\pi\ii}{3}}$.

\subsection{Untwisted and twisted correlators in the free matrix model}
\label{subsec:gaussian}
It is convenient to first consider what happens in the Gaussian multi-matrix model, {\it{i.e.}} when the interaction action $S_{\mathrm{int}}$ is neglected. All quantities computed in this free model
will be denoted with a subscript $0$.

As a consequence of the normalization of the integration measure (\ref{measure}), one easily sees that the partition function $\mathcal{Z}_0$ is 1, so that the vacuum expectation value of any function $f(a)$ is simply 
\begin{align}
	\label{gaussiandef}
	\big\langle f(a) \big\rangle_0 = \int \prod_{I=0}^{M-1} \Big( da_I\,\,\rme^{-\tr a_I^2}\Big)\,
	\, f(a)~. 
\end{align}	
Using the standard notation
\begin{align}
	\label{deftns}
	t_{n_1, n_2, \ldots} = \big\langle \tr a^{n_1}\, \tr a^{n_2} \ldots \big\rangle_0
\end{align}
for the expectation values in the Gaussian single-matrix model, we have for example
\begin{subequations}
\begin{align}
	\big\langle \tr a_{I}^n\, \tr a_{I}^m\big\rangle_0 &= t_{n,m}\qquad \forall ~I~,\label{keyu}\\[1mm]
	\big\langle \tr a_{I}^n \,\tr a_{J}^m\big\rangle_0 &= 
	\big\langle \tr a_{I}^n\big\rangle_0 ~\big\langle \tr a_{J}^m\big\rangle_0 
	= t_n \,t_m \qquad \forall~I\not= J~.\label{key}	
\end{align}
\label{keys}%
\end{subequations}

\subsubsection*{1-point correlators}
{From} the definitions (\ref{untA}) and (\ref{twal}), it is immediate to prove that
\begin{equation}
\big\langle A_{n} \big\rangle_0 =\frac{1}{\sqrt{M}}\,\sum_{I=0}^{M-1} \big\langle
\tr a_{I}^n \big\rangle_0= \sqrt{M}\,t_n~,
\label{1untw}
\end{equation}
and
\begin{equation}
\big\langle A_{\alpha,n} \big\rangle_0 =\frac{1}{\sqrt{M}}
\sum_{I=0}^{M-1} \, \rho^{-\alpha I}\,\big\langle \tr a_{I}^n \big\rangle_0 = 
\frac{t_n}{\sqrt{M}}\,\sum_{I=0}^{M-1}
 \, \rho^{-\alpha I}=0
\label{1tw}
\end{equation}
where the last step follows from the fact that
\begin{equation}
\sum_{I=0}^{M-1} \, \rho^{-\alpha I}=0
\label{sumroot}
\end{equation}
for any non-trivial $M$-th root of unity $\rho$ and any integer $\alpha\in [1,M-1]$.

\subsubsection*{2-point correlators}
In the free multi-matrix model, the correlators of two untwisted operators  (\ref{untA})  are
given by
\begin{align}
	\label{uu}
		\big\langle A_{n}\, A_{m}\big\rangle_0 
		= \frac{1}{M} \sum_{I,J=0}^{M-1} \big\langle \tr a_{I}^n \, \tr a_{J}^m\big\rangle_0
		= t_{n,m} + (M-1) \,t_n\, t_m~,
\end{align}
while those between untwisted and twisted operators vanish. Indeed, we have
\begin{align}
	\label{ut}
		\big\langle
		A_{n}\, A_{\alpha, m}\big\rangle_0 
		= \frac{1}{M} \sum_{I,J=0}^{M-1} \big\langle \tr a_{I}^n \, \tr a_{J}^m\big\rangle_0\,\,
		\rho^{-\alpha J} 
		= \frac{1}{M}\Big(t_{n,m} + (M-1) \,t_n\, t_m\Big)\sum_{J=0}^{M-1}\rho^{-\alpha J} =0~,
\end{align} 
where the last step follows from (\ref{sumroot}). Regarding the correlators of two twisted operators, 
these are non-zero only when the net $\mathbb{Z}_M$ charge vanishes modulo $M$; in fact 
\begin{align}
		\big\langle A_{\alpha, n}\, A_{\beta, m} \big\rangle_0 = \frac{1}{M} \sum_{I,J=0}^{M-1} \big\langle \tr a_{I}^n \, \tr a_{J}^m\big\rangle_0\,\,
		\rho^{-\alpha I-\beta J} =\frac{1}{M}\Big(t_{n,m} -\,t_n\, t_m\Big)\sum_{J=0}^{M-1}\rho^{-(\alpha +\beta)J} ~.
\end{align}
Using (\ref{sumroot}), this result can be expressed as
\begin{equation}
\big\langle A_{\alpha, n}\, A_{\beta, m} \big\rangle_0 =  t^c_{n,m}\,\delta_{\alpha+\beta,0}~,
\label{talphabeta}
\end{equation}
where the Kronecker $\delta$ is defined modulo $M$ and
\begin{align}
	\label{tcis}
		t^c_{n,m} = t_{n,m} - t_n\, t_m
\end{align}
is the connected correlator in the one-matrix model. Note that $t^c_{n,m}$ is different from
zero only when $(n+m)$ is even.
Furthermore, exploiting (\ref{realtau}), we can rewrite (\ref{talphabeta}) as
\begin{align}
	\label{conj2pt}
		\big\langle A_{\alpha, n}^{\phantom{\dagger}}\, A_{\alpha, m}^\dagger \big\rangle_0 =  t^c_{n,m}~.
\end{align}

The large-$N$ behavior of $t^c_{n,m}$ is described in Eq.s (3.8) and (3.9) of \cite{Beccaria:2020hgy}. As a consequence, in the multi-matrix model at large-$N$ we find
\begin{align}
	\label{Ndepcorr}
		\big\langle A_{\alpha, n}\, A_{\beta, m} \big\rangle_0 \,\propto\,  N^{\frac{n + m}{2}}\,\delta_{\alpha+\beta,0}
\end{align}
if $(n+m)$ is even. 

\subsubsection*{Higher-point correlators and Wick property}
The higher-point correlators can be discussed in a similar way. For brevity we just mention a few relevant features of the correlators involving only twisted operators, leaving aside the
cases of the untwisted or mixed correlators. 

For the 3-point correlators one finds that in the large-$N$ limit
\begin{align}
	\label{3ptcorr}
		\big\langle A_{\alpha, n}\, A_{\beta, m}\, A_{\gamma, p} \big\rangle_0 \,\propto
		\, N^{\frac{n+m+p}{2}-1}\, \delta_{\alpha+\beta+\gamma,0}
\end{align}
if $(n+m+p)$ is even; otherwise one gets zero. This result implies that in the large-$N$ limit
the factorization of a multi-point correlator in terms of 3-point correlators is always
suppressed  with respect to the factorization in terms of 2-point correlators.

The 4-point correlators at large-$N$ behave as follows 
\begin{align}
	\label{4ptcorr}
		\big\langle A_{\alpha, n}\, A_{\beta, m}\, A_{\gamma, p} \,A_{\delta,q} 
		\big\rangle_0 \,\propto
		\, N^{\frac{n+m+p+q}{2}}\, \delta_{\alpha+\beta+\gamma+\delta,0}
\end{align}
if $(n+m+p+q)$ is even. Sometimes it may happen, however, that the coefficient of the leading 
power of $N$ vanishes and the correlator scales as $N^{\frac{n+m+p+q}{2}-2}$. In this case the 4-point correlator is irreducible since it cannot be factorized in terms of 2-point correlators. When, instead, the leading term is as in (\ref{4ptcorr}), the correlator is reducible since the following
Wick-like decomposition 
\begin{align}
	\label{wick4}
		\big\langle A_{\alpha, n}\, A_{\beta, m}\, A_{\gamma, p}\, A_{\delta,q} \big\rangle_0 
		& =
		\big\langle A_{\alpha, n}\, A_{\beta, m}\big\rangle_0 \,
		\big\langle A_{\gamma, p}\, A_{\delta,q} \big\rangle_0 + 
		\big\langle A_{\alpha, n}\, A_{\gamma, p}\big\rangle_0 \,
		\big\langle A_{\beta, m}\, A_{\delta,q} \big\rangle_0 
		\notag\\[1mm]
	& \quad+
      \big\langle A_{\alpha, n}\, A_{\delta, q}\big\rangle_0 \,
      \big\langle A_{\beta, m}\, A_{\gamma,p} \big\rangle_0
\end{align}
holds at large $N$. In view of these facts, only the reducible 4-point correlators have to be considered in the large-$N$ limit, while the irreducible one, being subleading, can be neglected.

This pattern can be generalized to higher-point correlators without any difficulty.

\subsubsection*{Normal ordering}
The result  (\ref{talphabeta}) implies that, starting from $A_{\alpha,n}$, one can define the 
normal-ordered operators $P_{\alpha,n}$ in complete analogy to the one-matrix model by applying the Gram-Schimdt diagonalization procedure within each twisted sector. Therefore, at large-$N$ we have
\begin{align}
	\label{Ttotau}
		P_{\alpha, n} = A_{\alpha, n} + c_{n,n-2} \, A_{\alpha, n-2} + c_{n,n-4}  \,A_{\alpha, n- 4} + \ldots~,
\end{align}
where the coefficients $c_{n,r}$ are related to the power expansion of the Chebyshev polynomials \cite{Rodriguez-Gomez:2016ijh}. In closed form, we can write 
\begin{align}
	\label{Ttotauexpl}
		P_{\alpha, n} = n\, \sum_{k=0}^{\left[\frac{n-1}{2}\right]} (-1)^k \,\frac{N^k\,(n-k-1)!}{2^k \,k!\,(n-2k)!} \,A_{\alpha,n-2k}
\end{align}
with the understanding that $A_{\alpha,n}=0$ for $n=0,1$. 
The explicit expressions of $P_{\alpha,n}$ for the first few values of $n$ are
\begin{align}
	\label{ffT}
		P_{\alpha,2} & = A_{\alpha,2}~,\notag\\
		P_{\alpha,3} &= A_{\alpha,3}~,\notag\\
		P_{\alpha,4} & = A_{\alpha,4} - 2 N A_{\alpha,2}~,\notag\\
		P_{\alpha,5} &= A_{\alpha,5} - \frac 52  N A_{\alpha,3}~,\notag\\
		P_{\alpha,6} & = A_{\alpha,6} - 3  N A_{\alpha,4} + \frac 94 N^2 A_{\alpha,2}~,\notag\\
		P_{\alpha,7} &= A_{\alpha,7} - \frac 72  N A_{\alpha,5} + \frac 72 N^2 A_{\alpha,3}~.
\end{align}
The 2-point correlators of the operators $P_{\alpha,n}$ are diagonal by construction and read as follows:
\begin{align}
	\label{twoTT}
	 	\big\langle P_{\alpha,n}\, P_{\beta,m}\big\rangle_0 =  n \,\Big(\frac{N}{2}\Big)^n\,\delta_{n,m}\,\delta_{\alpha+\beta,0}~. 
\end{align}
We can further normalize these operators and define
\begin{align}
	\label{canT}
		\cP_{\alpha,n} = \frac{1}{\sqrt{n}} \, \Big(\frac{2}{N}\Big)^{\frac{n}{2}}\, P_{\alpha,n}
\end{align}
for which
\begin{align}
	\label{twoTTcan}
		\big\langle \cP_{\alpha,n}\,\cP_{\beta,m}\big\rangle_0 
		= \delta_{n,m}\,\delta_{\alpha+\beta,0}~. 
\end{align}
We can invert the relation (\ref{Ttotauexpl}) to express the operators $A_{\alpha,n}$ in terms 
of the normal-ordered ones $P_{\alpha,n}$ and then of the normalized operators $\cP_{\alpha,n}$, using the definition (\ref{canT}). Proceeding in this way, we explicitly find
\begin{align}
	\label{tautoThat}
		A_{\alpha,n} = \Big(\frac{N}{2}\Big)^ {\frac{n}{2}} 
		\,\sum_{k=0}^{\left[\frac{n-1}{2}\right]}
		\sqrt{n-2k} \,\binom{n}{k} \,\cP_{\alpha,n-2k}~.
\end{align}

This construction can of course be repeated also in the untwisted sector where one can define 
normal-ordered operators $P_n$ that are expressed in terms of the untwisted operators $A_n$ with a formula similar to (\ref{Ttotauexpl}), and satisfy analogous relations as the twisted
operators.

\subsection{The interaction action in terms of twisted operators}
\label{subsec:inttw}
Let us now consider the effect of interactions by taking into account the action (\ref{sinta}). Using the inverse of the map (\ref{twal}) one can show that 
\begin{align}
	\label{aIp1}
		\sum_{I=0}^{M-1} \big(\tr a_{I}^{n} - \tr a_{I+1}^{n}\big)\,
		 \big(\tr a_{I}^{m} - \tr a_{I+1}^{m}\big)
		= 
		 \sum_{\alpha_=1}^{M-1} 4\sin^2\big(\frac{\pi\alpha}{M}\big)\,
		A_{\alpha,n}^\dagger\, A_{\alpha,m}~.
\end{align}		
Since the factor $\sin^2\frac{\alpha\pi}{M}$  will frequently occur in the formulas below, we find convenient to introduce a name for it and thus we set
\begin{align}
	\label{salpha}
	\sin^2\big(\frac{\pi\alpha}{M}\big)\,\equiv \,s_\alpha ~.
\end{align}
The values of these factors for the first few values of $M$ are reported for convenience in Table~\ref{tab:sa}.
\begin{table}[ht]
	\begin{center}
	\setlength{\extrarowheight}{6pt}
	\begin{tabular}{c|ccccc}
		& $\alpha=1$ & $\alpha=2$ & $\alpha= 3$ & $\alpha = 4$ & $\alpha= 5$ \\[1mm]
		\hline
		$M=2$ & $1$ & & & & \\[1mm]
		$M=3$ & $\frac 34$ & $\frac 34$ & & & \\[1mm]
		$M=4$ & $\frac 12$ & $1$ & $\frac 12$ & & \\[1mm]
		$M=5$ & $\frac{5-\sqrt{5}}{8}$ & $\frac{5+\sqrt{5}}{8}$ & $\frac{5+\sqrt{5}}{8}$ & $\frac{5-\sqrt{5}}{8}$ & \\[1mm]
		$M=6$ & $\frac 14$ & $\frac 34$ & $1$ & $\frac 34$ & $\frac 14$ 
	\end{tabular}	
	\end{center}	
	\caption{The first few values of $s_\alpha =\sin^2\big(\frac{\pi\alpha}{M}\big)$.}
	\label{tab:sa}
\end{table}

The relation (\ref{aIp1}) implies a key fact: the interaction action (\ref{sinta}) becomes diagonal in the basis of twisted operators. In fact, one has
\begin{align}
	\label{Sinta}
		S_{\mathrm{int}}
		= \sum_{\alpha=1}^{M-1} \bigg[ 4 s_\alpha 
		\sum_{m=2}^\infty \sum_{k=2}^{2m}
		(-1)^{m+k} \Big(\frac{g^2}{8\pi^2}\Big)^{\!m} \,
		\binom{2m}{k}\,\frac{\zeta(2m-1)}{2m} 
		\, A_{\alpha,2m-k}^\dagger \,A_{\alpha,k}\bigg]~.
\end{align}
We can rewrite the right-hand side in terms of the normalized normal-ordered operators
$\cP_{\alpha,n}$ by means of (\ref{tautoThat}). Rearranging the sums, we can obtain a compact form if we introduce the infinite column vectors 
$\bcP_\alpha$ containing all operators in the twisted class $\alpha$:
\begin{align}
	\label{defPvec}
		\bcP_{\alpha}= \begin{pmatrix}
		\cP_{\alpha,2}\\
		\cP_{\alpha,3} \\
		\cP_{\alpha,4} \\
		\vdots
		\end{pmatrix}~. 
\end{align}
In fact, after some straightforward manipulations, we get
\begin{align}
	\label{SintX}
		S_{\mathrm{int}} = - \frac{1}{2} \sum_{\alpha=1}^{M-1} 
		s_\alpha\,\bcP_\alpha^\dagger \,\Xx\, \bcP_\alpha
\end{align}
where $\Xx$ is an infinite symmetric matrix whose elements $\Xx_{r,s}$ vanish if $r$ and $s$ have opposite parity, namely
\begin{align}
	\label{Xeo0}
		\Xx_{2k+1,2\ell} = 0~,
\end{align}
while, if they are both even or both odd, they are given by
\begin{align}
	\label{Xis}
		\Xx_{r,s} = -8 \sqrt{r\, s} \,\sum_{p=0}^\infty (-1)^p \,c_{r,s,p}\, 
		\frac{\zeta(r+s+2p-1)}{r+s+2p}\, 
		\Big(\frac{\lambda}{16\pi^2}\Big)^{\frac{r+s+2p}{2}}~,
\end{align}
with $\lambda$ being the 't Hooft coupling (\ref{lambda}) and
\begin{align}
	\label{crsp}
		c_{r,s,p} = \sum_{\ell=0}^p \frac{(r+s+2p)!}{\ell!\,(p-\ell)!\,(r+\ell)!\,(s+p-\ell)!}
		= \frac{(r+s+2p)!\,(r+s+2p)!}{p!\,(r+p)!\,(s+p)!\,(r+s+p)!}~.
\end{align}

For the following computations we find convenient to separate the odd and even entries of the matrix $\Xx$ writing
\begin{align}
	\label{Xoddeven}
		\Xx^{\mathrm{odd}}_{k,\ell}\, \equiv\, \Xx_{2k+1,2\ell+1}\qquad\mbox{and}\qquad
		\Xx^{\mathrm{even}}_{k,\ell} \,\equiv\, \Xx_{2k,2\ell}~.
\end{align} 
These entries can be expressed in terms of integrals of Bessel functions of the first kind $J_n$. In fact one has
\begin{align}
	\label{Xoddis}
		\mathsf{X}^{\mathrm{odd}}_{k,\ell}&=
		-8 (-1)^{k+\ell} \sqrt{(2k+1)(2\ell+1)} \int_0^\infty \!\frac{dt}{t}\, 
		\frac{\rme^t}{(\rme^t-1)^2}\,
		J_{2k+1}\Big(\frac{t\sqrt{\lambda}}{2\pi}\Big)\, 
		J_{2\ell+1}\Big(\frac{t\sqrt{\lambda}}{2\pi}\Big)~.
\end{align}
This is the same expression originally introduced in \cite{Beccaria:2020hgy} to write the interaction action of the matrix model for the so-called $\mathbf{E}$ theory\,%
\footnote{As we will see in Appendix \ref{app:Z2}, this fact is not a coincidence.}. 
Similarly, one finds
\begin{align}
	\label{Xevenis}
		\mathsf{X}^{\mathrm{even}}_{k,\ell}&=
		-8 (-1)^{k+\ell} \sqrt{(2k)(2\ell)} \int_0^\infty \!\frac{dt}{t}\, 
		\frac{\rme^t}{(\rme^t-1)^2}\,
		J_{2k}\Big(\frac{t\sqrt{\lambda}}{2\pi}\Big)\, 
		J_{2\ell}\Big(\frac{t\sqrt{\lambda}}{2\pi}\Big)~.
\end{align}
Notice that the perturbative expression (\ref{Xis}) is recovered by Taylor expanding the Bessel functions for small $\lambda$ and then performing the resulting integral over $t$. 
However, the advantage of using the integral representations (\ref{Xoddis}) and (\ref{Xevenis}) is that
they can be also expanded asymptotically in the regime of large $\lambda$ using the inverse Mellin
transform of the product of two Bessel functions. This is what we will discuss in Section~\ref{sec:exact}.

Let us finally remark that, while (\ref{SintX}) expresses the interaction action in a very compact way, the contributions from the $\alpha$-th and $(M-\alpha)$-th sectors are equal because 
$\cP_{\alpha}^\dagger = \cP_{M-\alpha}$. Thus, we can also write
\begin{align}
	\label{SintXbis}
	S_{\mathrm{int}} =  \sum_{\alpha=1}^{\left[\frac{M-1}{2}\right]}  
	S_{\mathrm{int}}^{(\alpha)} + S_{\mathrm{int}}^{(M/2)}  ~,
\end{align}
with
\begin{align}
	\label{keyS}	
		S_{\mathrm{int}}^{(\alpha)} = - s_\alpha\, \bcP_\alpha^\dagger\, \Xx\, \bcP_\alpha
		\qquad\mbox{and}\qquad
			S_{\mathrm{int}}^{(M/2)} = - \frac{1}{2} \, \bcP_{M/2}^T \,\Xx\, \bcP_{M/2}~.		  
\end{align}
Of course the last term is present only when $M$ is even.

The above results make explicit the fact that, like in the Gaussian model, also in the interacting theory
the correlators of the quiver multi-matrix model factorize completely in the various twisted sectors. Therefore, within each twisted sector $\alpha$, the vacuum expectation value
of any function of $\cP_{\alpha,n}$ is given by
\begin{equation}
\big\langle f(\cP_{\alpha,n})\big\rangle=
\,\frac{\displaystyle{ \int \prod_{I=0}^{M-1}\Big( da_I\,\,\rme^{-\tr a_I^2}\Big) \,f(\cP_{\alpha,n})\,
\rme^{-S_{\mathrm{int}}^{(\alpha)}}}}
		{\displaystyle{ \int \prod_{I=0}^{M-1}\Big( da_I\,\,\rme^{-\tr a_I^2}\Big) \,
\rme^{-S_{\mathrm{int}}^{(\alpha)}}}}
=\frac{\big\langle \,f(\cP_{\alpha,n}) \,\rme^{-S_{\mathrm{int}}^{(\alpha)}}\,\big\rangle_0\phantom{\Big|}}{\big\langle \,\rme^{-S_{\mathrm{int}}^{(\alpha)}}\,\big\rangle_0\phantom{\Big|}}~.
\label{vevalpha}
\end{equation}
In this approach everything is reduced to the calculation of vacuum expectation values in the free
matrix model, and these can be computed using the formulas presented in Section~\ref{subsec:gaussian}.

\section{Exact results at large $N$}
\label{sec:exact}
In the large-$N$ limit the correlators of twisted operators can be given an exact formal 
expression in terms of the matrix $\Xx$ introduced in the previous section. In view of the Wick property of the multi-correlators, we focus on the 2-point functions and show how 
to extract their perturbative expansions up to very high order in $\lambda$ with great efficiency.
These expansions can then be resummed \`a la Pad\'e and successfully compared with a numerical evaluation based on a Monte Carlo simulation. Moreover, using
the integral representation of the $\Xx$ matrix, we can obtain also the leading behavior
of the twisted correlators for large values of the 't Hooft coupling and provide an effective strong-coupling description in terms of a generating function.

To derive these results, in view of the remarks made at the end of the previous section, we can work independently within each twisted sector, in which the computations amount to a rather 
straightforward adaptation of the techniques employed in \cite{Beccaria:2020hgy,Beccaria:2021hvt}.

\subsection{The effective Gaussian formulation}
\label{subsecn:gaussian}
We find convenient to start from the free matrix model in which the operators $\cP_{\alpha,n}$ defined in (\ref{canT}) enjoy, at large $N$, the Wick property with respect to the propagator
(\ref{twoTTcan}) that we rewrite as
\begin{equation}
\label{corrfact}
\big\langle \cP_{\alpha,n}^{\phantom{\dagger}} \,\cP_{\alpha,m}^\dagger\big\rangle_0 = \delta_{n,m}
\end{equation}
for each twisted sector $\alpha$. The correlators of several operators $\cP_{\alpha,n}$ can then be rephrased in terms of Gaussian integrals over ordinary complex variables $z_{\alpha,n}$ by
writing
\begin{align}
	\label{gausseta}
		\big\langle
		\cP_{\alpha,n_1}^{\phantom{\dagger}} \,\cP_{\alpha,n_2}^\dagger \,
		\cP_{\alpha,n_3}^{\phantom{\dagger}} \,\cP_{\alpha,n_4}^\dagger \ldots\big\rangle_0 = 
		\int \!d^2\boldsymbol{z}_\alpha~ z_{\alpha,n_1}\, z_{\alpha,n_2}^* \,z_{\alpha,n_3}\, z_{\alpha,n_4}^* \ldots \rme^{-\boldsymbol{z}_\alpha^\dagger \boldsymbol{z}_\alpha}~.
\end{align}		  
In the right-hand side we have introduced the infinite vector
\begin{align}
	\label{zvec}
		\boldsymbol{z}_\alpha
		= \begin{pmatrix}
		z_{\alpha,2}\\
		z_{\alpha,3}\\
		\vdots
\end{pmatrix}		 
\end{align}
and defined the integration measure
\begin{align}
	\label{d2z}
		d^2\boldsymbol{z}_\alpha = \prod_{n=2}^\infty \frac{d^2 z_{\alpha,n}}{\pi}	
\end{align}
in such a way that
\begin{align}
	\label{gaussz}
		\int\! d^2\boldsymbol{z}_\alpha~ \rme^{- \boldsymbol{z}_\alpha^\dagger \boldsymbol{z}_\alpha} = 1~.
\end{align}	
With these conventions, one has the ``propagator''
\begin{align}
	\label{zprop}
		\int \!d^2\boldsymbol{z}_\alpha~ z_{\alpha,n}^* z_{\alpha,m} \,
		\rme^{- \boldsymbol{z}_\alpha^\dagger \boldsymbol{z}_\alpha} = \delta_{n,m}
\end{align}
which is the counterpart of (\ref{corrfact}).
Notice that in the twisted sector with $\alpha=M/2$, which occurs when $M$ is even, 
the operators  $\cP_{M/2,n}\equiv \cP_n$ are real, and thus their correlators are expressed 
in terms of integrals over real variables $z_{n}$ according to
\begin{align}
	\label{gaussetareal}
		\big\langle \cP_{n_1} \cP_{n_2} \ldots\big\rangle_0 = 
		\int \!d\boldsymbol{z}~ z_{n_1} z_{n_2}  \ldots \,
		\rme^{-\frac 12 \boldsymbol{z}^T \boldsymbol{z}}~.
\end{align}		  

Let us now consider the interacting matrix model corresponding to the $\cN=2$ quiver theory.
As we have seen in Section~\ref{subsec:inttw}, within a generic twisted sector $\alpha$
the presence of the interaction amounts to the insertion of the factor
\begin{equation}
\rme^{-S_{\mathrm{int}}^{(\alpha)}}= \rme^{s_\alpha\, \bcP_\alpha^\dagger\, \Xx\, \bcP_\alpha}
\end{equation}
in the Gaussian model. Then, taking into account that the Wick rule exponentiates, the partition
function $\cZ_\alpha$ can be written as
\begin{align}
	\label{Zalphaz}
		\cZ_\alpha & = \big\langle \rme^{s_\alpha\, \bcP_\alpha^\dagger\, \Xx\, \bcP_\alpha} \big\rangle_0
		= \int \! d^2\boldsymbol{z}_\alpha ~
		\rme^{- \boldsymbol{z}_\alpha^\dagger \boldsymbol{z}_\alpha}\,\,
		\rme^{s_\alpha\, \boldsymbol{z}_\alpha^\dagger\, \Xx \,\boldsymbol{z}_\alpha}~,
\end{align}
from which we get\,%
\footnote{Let us remark that the matrix $\Xx$ would be block-diagonal if we reordered the basis separating the even and odd entries according to (\ref{Xeo0}) and (\ref{Xoddeven}). 
In this way we have 
	\begin{align*}
	\det \big(\mathbb{1} - s_\alpha \Xx\,\big) = 
	\det \big(\mathbb{1} - s_\alpha \Xx^{\rm even}\,\big) \times
	\det \big(\mathbb{1} - s_\alpha \Xx^{\rm odd}\,\big)~.
	\end{align*}
} 
\begin{align}
	\label{Zalphazres}
		\cZ_\alpha = \Big[\det \big(\mathbb{1} - s_\alpha \,\Xx\,\big)\Big]^{-1}~.	
\end{align}	
Analogously, for $\alpha=M/2$ we have	
\begin{align}
	\label{Zalphazresm2}
		\cZ_{M/2} = \Big[\det \big(\mathbb{1} -  \Xx\,\big)\Big]^{-\frac 12}~.	
\end{align}	
In this formulation one can easily see that the expectation values (\ref{vevalpha})
become
\begin{align}
	\label{vevPz}
		\big\langle f(\cP_{\alpha,n})\big\rangle = \frac{1}{Z_\alpha} \int\! 
		d^2\boldsymbol{z}_\alpha~f(z_{\alpha,n}) \,\,
		\rme^{- \boldsymbol{z}_\alpha^\dagger \,\big(\mathbb{1} - s_\alpha \,\Xx\,\big)\,\boldsymbol{z}_\alpha}~,
\end{align}
from which it follows that
\begin{align}
	\label{tpP}
		\big\langle \cP_{\alpha,n}^{\phantom{\dagger}}\, \cP_{\alpha,m}^\dagger\big\rangle = 
		\Big(\frac{1}{\,\mathbb{1} - s_\alpha\,\Xx\,}\Big)_{n,m}\,
		\equiv \,\big(\Dx_\alpha\big)_{n,m}~.
\end{align}
For $\alpha= M/2$, using (\ref{gaussetareal}) and (\ref{Zalphazresm2}), we find
\begin{align}
	\label{tpPm2}
		\big\langle \cP_{n} \,\cP_{m}\big\rangle = 
		\Big(\frac{1}{\,\mathbb{1} - \Xx\,}\Big)_{n,m} \,
		\equiv \, \big(\Dx_{M/2} \big)_{n,m}
\end{align}
which is exactly what one would get by extending (\ref{tpP}) to $\alpha=M/2$, 
since $s_{M/2}= \sin^2(\pi/2)= 1$. Thus, from now on, we will not distinguish this case any longer and write formulas that apply to all twisted classes, including $\alpha=M/2$.

To make (\ref{tpP}) and (\ref{tpPm2}) more explicit, it is convenient to use the matrices $\Xx^{\rm even}$ and $\Xx^{\rm odd}$ introduced in (\ref{Xoddeven}), in terms of which we have
\begin{align}
	\label{tpPeven}
		\big\langle \cP_{\alpha,2k}^{\phantom{\dagger}}\,\cP_{\alpha,2\ell}^\dagger \big\rangle = 
		\Big(\frac{1}{\,\mathbb{1} - s_\alpha\, \Xx^{\mathrm{even}}\,}\Big)_{k,\ell} \,
		\equiv \, \big(\Dx^{\mathrm{even}}_\alpha\big)_{k,\ell}
\end{align}
and 
\begin{align}
	\label{tpPodd}
		\big\langle \cP_{\alpha,2k+1}^{\phantom{\dagger}}\,\cP_{\alpha,2\ell+1}^\dagger\big\rangle = 
		\Big(\frac{1}{\,\mathbb{1} - s_\alpha\, \Xx^{\mathrm{odd}}\,}\Big)_{k,\ell} \,
		\equiv \, \big(\Dx^{\mathrm{odd}}_\alpha\big)_{k,\ell}~.
\end{align}
Correlators involving multiple pairs of $\cP_{\alpha,n}\,\cP_{\alpha,m}^\dagger$
are computed using the Wick rule with the propagator $\Dx_\alpha$.

We point out that in the above expressions the $\lambda$-dependence is \emph{entirely} 
encoded in the matrices $\Xx^{\mathrm{even}}\,$ and $\,\Xx^{\mathrm{odd}}$, and thus 
the formulas (\ref{tpPeven}) and (\ref{tpPodd}) are \emph{exact} in $\lambda$.

\subsection{Twisted correlators of normal ordered operators}
\label{subsec:twistedcorr}

The twisted operators $\cP_{\alpha,n}$ which were mutually orthogonal in the Gaussian model, are no longer so in the interacting matrix model and thus they cannot represent the twisted operators
$T_{\alpha,n}(\vec{x})$ of the quiver theory, which are normal-ordered and mutually orthogonal
(see (\ref{TT})). This problem is easily cured by 
applying the Gram-Schmidt procedure to the space of the operators $\cP_{\alpha,n}$ 
with scalar product $(\Dx_\alpha)_{nm}$; in this way we obtain a set of operators $\cT_{\alpha,n}$
which are mutually orthogonal in the interacting theory and thus can represent the twisted operators $T_{\alpha,n}(\vec x)$, normalized so as to have a unit correlator in the free theory. 
In particular, the 2-point functions of the matrix operators $\cT_{\alpha,n}$ capture, in the
large-$N$ limit, the ratio of the 2-point functions in the quiver gauge theory to the ones in the non interacting theory. Since the latter coincide with the 2-point functions in the $\cN=4$ SU($N$) SYM theory, we can write
\begin{align}
	\label{cTT}
	\frac{\big\langle T_{\alpha,n}^{\phantom{\dagger}}(\vec{x}) \,T_{\alpha,n}^{\,\dagger}(\vec{0})\big\rangle\phantom{\Big|}}{~\big\langle 
O_{n}(\vec{x})\, \overbar{O}_{n}(\vec{0})\big\rangle_{0}\phantom{\Big{|}}}=
		\big\langle 
		\cT_{\alpha,n}^{\phantom{\dagger}}\,\cT_{\alpha,n}^{\,\dagger} \big\rangle
		 + \ldots		
\end{align}
where the correlator in the right-hand side takes the form
\begin{align}
	\label{tpDelta}		
		\big\langle 
		\cT_{\alpha,n}^{\phantom{\dagger}}\,\cT_{\alpha,n}^{\,\dagger} \big\rangle
		=  1 + \Delta_{\alpha,n}(\lambda)
\end{align}
with $\Delta_{\alpha,n}(\lambda)$ vanishing for $\lambda\to 0$, and the
ellipses mean $O(1/N^2)$ corrections.

The ratio in (\ref{cTT}) is precisely what we have computed using Feynman diagrams in Section~\ref{secn:diagrams} at the first orders in perturbation theory. The matrix model approach
provides an \emph{exact} expression for this ratio in which the dependence on the coupling
constant is completely determined. Indeed, the Gram-Schmidt procedure explicitly constructs 
the twisted operators $\cT_{\alpha,n}$ in the form
\begin{equation}
\cT_{\alpha,n} = \cP_{\alpha,n} + \ldots 
\end{equation}
where the dots stand for a combination of operators $\cP_{\alpha,k}$ with $k<n$, devised in such a way that $\cT_{\alpha,n}$ is orthogonal to all the operators of lower dimension. The combination with this property is obtained in terms of the matrix $\big(\Dx_\alpha\big)_{n,m}$ introduced
above, which is non-zero only if $n$ and $m$ have the same parity. This means that we actually
carry out two separate Gram-Schmidt procedures for the even and odd operators, 
based on the use of $\Dx_\alpha^{\mathrm{even}}$ and $\Dx_\alpha^{\mathrm{odd}}$ respectively.
What is most relevant is that these Gram-Schmidt procedures also give an explicit expression of the 2-point functions (\ref{tpDelta}) since one has
\begin{subequations}
\begin{align}
		\big\langle\cT_{\alpha,2k}^{\phantom{\dagger}}\,\cT_{\alpha,2k}^{\,\dagger} \big\rangle&=1 + \Delta_{\alpha,2k}(\lambda)= 
		\frac{\det\big[(\Dx_\alpha^{\mathrm{even}})_{(k)}\big]\phantom{\Big|}}{\det\big[(\Dx_\alpha^{\mathrm{even}})_{(k-1)}\big]\phantom{\Big|}}~,\label{gammaome0even}\\[1mm]
		\big\langle\cT_{\alpha,2k+1}^{\phantom{\dagger}}\,\cT_{\alpha,2k+1}^{\,\dagger} \big\rangle&=
		1 + \Delta_{\alpha,2k+1}(\lambda) = 
		\frac{\det\big[(\Dx_\alpha^{\mathrm{odd}})_{(k)}\big]\phantom{\Big|}}{\det\big[(\Dx_\alpha^{\mathrm{odd}})_{(k-1)}\big]\phantom{\Big|}}\label{gammaome0odd}
\end{align}
\label{gammaome0evod}%
\end{subequations}
Here the notation $M_{(k)}$ indicates the upper-left $k\times k$ block of a matrix $M$, with the convention that $M_{(0)}= 1$. The ratios of determinants in (\ref{gammaome0evod}) can also be rewritten in an even more explicit form as explained in \cite{Beccaria:2020hgy}. Indeed, if we denote 
by $M_{[k]}$ the sub-matrix obtained from $M$ by removing its first $(k-1)$ rows and columns,
with the convention that $M_{[1]} = M$, then from the definition (\ref{tpP}) we can prove that
\begin{subequations}
\begin{align}
		1 + \Delta_{\alpha,2k}(\lambda) &=
		\bigg(\frac{1}{\,\mathbb{1} - s_\alpha\, \Xx^{\mathrm{even}}_{[k]}\,}\bigg)_{1,1} ~,\\[1mm]
		1 + \Delta_{\alpha,2k+1}(\lambda) &=
		\bigg(\frac{1}{\,\mathbb{1} - s_\alpha\, \Xx^{\mathrm{odd}}_{[k]}\,}\bigg)_{1,1} ~.
\end{align}
\label{gammaome}%
\end{subequations}
for any integer $k\geq 1$. We stress once again that these expressions are \emph{exact} in
$\lambda$ since the full dependence on the coupling constant is captured by the matrices
$\Xx^{\mathrm{even}}_{[k]}$ and $\Xx^{\mathrm{odd}}_{[k]}$.

When $\lambda\to 0$, we can easily expand the previous expressions and find the perturbative series
in a very efficient way. Indeed, one first writes
\begin{subequations}
\begin{align}
\Delta_{\alpha,2k}(\lambda) &=s_\alpha \Big(\Xx^{\mathrm{even}}_{[k]}\Big)_{1,1} 
+ s_\alpha^2 \Big(\big(\Xx^{\mathrm{even}}_{[k]}\big)^2\Big)_{1,1} 
+ s_\alpha^3 \Big(\big(\Xx^{\mathrm{even}}_{[k]}\big)^3\Big)_{1,1}+\ldots~,\\[1mm]
\Delta_{\alpha,2k+1}(\lambda) &=s_\alpha \Big(\Xx^{\mathrm{odd}}_{[k]}\Big)_{1,1} 
+ s_\alpha^2 \Big(\big(\Xx^{\mathrm{odd}}_{[k]}\big)^2\Big)_{1,1} 
+ s_\alpha^3 \Big(\big(\Xx^{\mathrm{odd}}_{[k]}\big)^3\Big)_{1,1}+\ldots~,
\end{align}
\label{Deltaexp}%
\end{subequations}
and then exploits the integral representations (\ref{Xevenis}) and (\ref{Xoddis}) in terms of
Bessel functions and the sum rules that the latter satisfy, in order to obtain the Taylor expansion
of the right-hand sides of (\ref{Deltaexp}) for small $\lambda$.
For instance, up to order $\lambda^7$ and for the first few values of $k$ we explicitly get
\begin{subequations}
\begin{align}
		\Delta_{\alpha,2}(\lambda)  &= -\frac{3\,s_\alpha\,\zeta(3)}{32\,\pi^4}\lambda^2+\frac{5\,s_\alpha\, \zeta(5)}{64\,\pi^6}\lambda^3-
		\Big(\frac{245\,s_\alpha\, \zeta (7)}{4096\,\pi^{8}}-\frac{9\,s_\alpha^2\,\zeta (3)^2}{1024\,\pi^{8}}\Big)\lambda^4\notag\\
		&~\quad+\Big(\frac{189\,s_\alpha\, \zeta (9)}{4096\,\pi^{10}}-\frac{15\,s_\alpha^2\,\zeta (3)\zeta(5)}{1024\,\pi^{10}}\Big)\lambda^5\notag\\
		&~\quad-\Big(\frac{38115\,s_\alpha\, \zeta (11)}{1048576\,\pi^{12}}-\frac{735\,s_\alpha^2\,\zeta (3)\zeta(7)}{65536\,\pi^{12}}-\frac{825\,s_\alpha^2\,\zeta (5)^2}{131072\,\pi^{12}}
		+
		\frac{27\,s_\alpha^3\,\zeta (3)^3}{32768\,\pi^{12}}\Big)\lambda^6\label{ffoD2}\\
		&~\quad
+		\Big(\frac{61347\,s_\alpha\, \zeta (13)}{2097152\,\pi^{14}}-\frac{567\,s_\alpha^2\,\zeta (3)\zeta(9)}{65536\,\pi^{14}}-\frac{1295\,s_\alpha^2\,\zeta (5)\zeta(7)}{131072\,\pi^{14}}
+\frac{135\,s_\alpha^3\, \zeta (3)^2\zeta(5)}{65536\,\pi^{14}}\Big)\lambda^7
+\ldots~,\notag\\[4mm]
		\Delta_{\alpha,3}(\lambda) & =-\frac{5\,s_\alpha\,\zeta(5)}{256\,\pi^6}\lambda^3+\frac{105\,s_\alpha\, \zeta(7)}{4096\,\pi^8}\lambda^4-\frac{1701\,s_\alpha\, \zeta (9)}{65536\,\pi^{10}} \lambda^5+
		\Big(\frac{12705\,s_\alpha\, \zeta (11)}{524288\,\pi^{12}}+\frac{25\,s_\alpha^2\,\zeta (5)^2}{65536\,\pi^{12}}\Big)\lambda^6\notag\\
		&~\quad-\Big(\frac{184041\,s_\alpha\, \zeta (13)}{8388608\,\pi^{14}}+\frac{525 \,s_\alpha^2\,\zeta (5) \zeta (7)}{524288\,\pi^{14}}\Big)\lambda^7+\ldots~,	\label{ffoD3}\\[4mm]
		\Delta_{\alpha,4}(\lambda) &= -\frac{35\,s_\alpha\,\zeta(7)}{8192\pi^{8}}\lambda^4+\frac{63\,s_\alpha\,\zeta(9)}{8192\pi^{10}}\lambda^5-\frac{2541\,s_\alpha\,\zeta(11)}{262144\pi^{12}}\lambda^6+\frac{5577\,s_\alpha\,\zeta(13)}{524288\pi^{14}}\lambda^7+\ldots~,\label{ffoD4}\\[4mm]
		\Delta_{\alpha,5}(\lambda) &= -\frac{63\,s_\alpha\,\zeta(9)}{65536\pi^{10}}\lambda^5+\frac{1155\,s_\alpha\,\zeta(11)}{524288\pi^{12}}\lambda^6-\frac{27885\,s_\alpha\,\zeta(13)}{8388608\pi^{14}}\lambda^7+\ldots~.\label{ffoD5}
\end{align}
\label{explicitDelta}%
\end{subequations}
One can easily verify that the first perturbative terms in these expressions
precisely match the results obtained from Feynman diagrams in Section~\ref{secn:diagrams} for the quiver theories with $M=2,3,4$. This comparison also shows that the overall numerical coefficients
that we pointed out at the end of Section~\ref{secn:diagrams} are precisely the values of the
coefficients $s_\alpha$ appropriate for the twisted sectors in the different theories.

This procedure can be efficiently implemented in a computer code; this allows us to push these expansions to very high orders with little effort. The resulting perturbative series have a finite convergence radius and are valid up to $\lambda = \pi^2$ \cite{Beccaria:2020hgy,Beccaria:2021vuc,Beccaria:2021hvt}; however, they can be taken as an input for a Pad\'e resummation and the resulting resummed expressions can be extended to values of $\lambda$ well beyond the perturbative bound, and compared with numerical simulations at intermediate or strong coupling.

\subsection{Strong-coupling behavior}
\label{subsec:sc}

The integral representations (\ref{Xoddis}) and (\ref{Xevenis}) of the infinite matrices 
$\Xx^{\mathrm{odd}}$ and $\Xx^{\mathrm{even}}$ in terms of products of two Bessel functions allows us to obtain their strong-coupling behavior. Indeed, as shown in \cite{Beccaria:2021vuc,Beccaria:2021hvt}, writing the product of the Bessel functions as an 
inverse Mellin transform, one finds that
\begin{align}
	\label{xsodd}
		\Xx^{\mathrm{odd}} \underset{\lambda \to \infty}{\sim}
		- \frac{\lambda}{2 \pi^2}  \, \Sx^{\mathrm{odd}} ~,
\end{align}
where $\Sx^{\rm odd}$ is a three-diagonal infinite matrix of elements
\begin{align}
	\label{sodd}
		\big(\Sx^{\mathrm{odd}}\big)_{k,\ell}
		= \sqrt{\frac{2\ell+1}{2k+1}} \,
		\Big(\!-\frac{\delta_{k-1,\ell}}{2\, (2k)\,(2k-1)} + \frac{\delta_{k,\ell}}{(2k)\,(2k+2)} -
		\frac{\delta_{k+1,\ell}}{2\,(2k+2)\,(2k+3)}\Big) ~.
\end{align}
In Appendix \ref{app:b}, we show that an analogous result holds in the even case, namely
\begin{align}
	\label{xseven}
		\Xx^{\mathrm{even}} \underset{\lambda \to \infty}{\sim}
		- \frac{\lambda}{2 \pi^2}  \, \Sx^{\mathrm{even}} ~,
\end{align}
with 
\begin{align}
	\label{seven}
		\big(\Sx^{\mathrm{even}}\big)_{k,\ell}
		= \sqrt{\frac{\ell}{k}} \, 
		\Big(\!- \frac{\delta_{k-1,\ell}}{2 (2k-2) (2k-1)} + \frac{\delta_{k,\ell}}{(2k-1)(2k+1)} 
		-\frac{\delta_{k+1,\ell}}{2(2k+1)(2k+2)}
		\Big)~. 
\end{align}
Therefore, from (\ref{gammaome}) we find that at strong coupling
\begin{subequations}
\begin{align}
		1 + \Delta_{\alpha,2k+1}(\lambda) &\,\underset{\lambda \to \infty}{\sim}\,
		 \frac{2\pi^2}{\lambda\,s_\alpha} \Big( \big(S^{\mathrm{odd}}_{[k]}\big)^{-1}\Big)_{1,1}~,\\[1mm]
		1 + \Delta_{\alpha,2k}(\lambda) &\,\underset{\lambda \to \infty}{\sim}\,
		 \frac{2\pi^2}{\lambda\,s_\alpha} \Big( \big(S^{\mathrm{even}}_{[k]}\big)^{-1}\Big)_{1,1}~.
\end{align}
\label{DelS}%
\end{subequations}
In \cite{Beccaria:2021hvt} it was analytically proven that
\begin{align}
	\label{Sodd11}
		\Big( \big(S^{\mathrm{odd}}_{[k]}\big)^{-1}\Big)_{1,1}
		= \frac{\det S^{\mathrm{odd}}_{[k+1]}\phantom{\Big|}}{\det S^{\mathrm{odd}}_{[k]}\phantom{\Big|}}
		= 2(2k+1)(2k)~.
\end{align}
Analogously, in Appendix \ref{app:b} we show that
\begin{align}
	\label{Seven11}
\Big( \big(S^{\mathrm{even}}_{[k]}\big)^{-1}\Big)_{1,1}
		= \frac{\det S^{\mathrm{even}}_{[k+1]}\phantom{\Big|}}{\det S^{\mathrm{even}}_{[k]}\phantom{\Big|}}
		= 2(2k)(2k-1)~.
\end{align}
Inserting these results into (\ref{DelS}), we easily realize that the odd and even cases can be compactly written in a single formula as follows
\begin{align}
	\label{Deloddeven}
	1 + \Delta_{\alpha,n}(\lambda) \,\underset{\lambda \to \infty}{\sim}\, \frac{4\pi^2}{\lambda\,s_\alpha} n(n-1)~.
\end{align}
Thus, in the planar limit the leading strong coupling behavior of the 2-point functions 
of the normal-ordered operators in the matrix model is
\begin{align}
	\label{TTfin}
		\big\langle \cT_{\alpha,n}^{\phantom{\dagger}}\, \cT_{\alpha,n}^{\,\dagger}
		\big\rangle \,\underset{\lambda \to \infty}{\sim}\,\frac{4\pi^2}{\lambda\,s_\alpha} n(n-1)~.
\end{align}
According to (\ref{cTT}), this is also the strong-coupling behavior of the 2-point functions
of the twisted primary operators in the quiver gauge theory at large $N$ 
(normalized to the ones of the $\cN=4$ SYM theory).

\subsection{Description through  a generating function}
We can summarize our findings by saying that in the planar limit the correlators of the normal-ordered twisted operators $\cT_{\alpha,n}$ are obtained from the Wick rule with the propagator (\ref{tpDelta}) which at strong 't Hooft coupling behaves asymptotically as in (\ref{TTfin}).
We now rephrase these results by introducing for each twisted sector $\alpha$
a set of complex variables $\eta_{\alpha,n}$, collected in an infinite vector $\boldsymbol{\eta}_\alpha$, that play the role of sources for the operators 
$\cT_{\alpha,n}$. Let us denote the corresponding generating function by
\begin{align}
	\label{genZ}
		Z_\alpha[\boldsymbol{\eta}_\alpha] = \big\langle
		\rme^{\,\boldsymbol{\eta}^{\,\dagger}_\alpha \,\boldsymbol{\cT}_\alpha 
		+ \boldsymbol{\cT}_\alpha^{\,\dagger}\, \boldsymbol{\eta}_\alpha}\big\rangle~.
\end{align}	
This expression generates multiple correlators through the relation
\begin{align}
	\label{correta1}
		\big\langle
		\cT_{\alpha,n}^{\phantom{\dagger}}\,\cT_{\alpha,m}^{\,\dagger} \,\cT_{\alpha,p}^{\phantom{\dagger}}\,\cT_{\alpha,q}^{\,\dagger} \ldots\big\rangle
		= \frac{\,~\partial~\,}{\partial \eta_{\alpha,n}^{\,\dagger}} \, \frac{\partial}{\partial \eta_{\alpha,m}^{\phantom{\dagger}}} \,
		\frac{\,~\partial~\,}{\partial \eta_{\alpha,p}^{\,\dagger}}
		\, \frac{\partial}{\partial \eta_{\alpha,q}^{\phantom{\dagger}}}\ldots
		Z_\alpha[\boldsymbol{\eta}_\alpha]\,\bigg|_{\boldsymbol{\eta}_\alpha=0}~.	
\end{align}
Our results amount to the statement that at large $N$ the generating function 
$Z_\alpha[\boldsymbol{\eta}_\alpha]$ becomes Gaussian and reads
\begin{align}
	\label{Zetaetais}
		Z_\alpha[\boldsymbol{\eta}_\alpha] = \exp\bigg[\,\sum_{n=2}^\infty 
		\eta_{\alpha,n}^\dagger \,\big(1 + \Delta_{\alpha,n}(\lambda)\big)\,
		\eta_{\alpha,n}\bigg]~.
\end{align}
Thus we can introduce a quadratic effective action
\begin{align}
	\label{Seffis}
		S_{\alpha}^{\rm eff}[\boldsymbol{\eta}_\alpha] = -\log Z_\alpha[\boldsymbol{\eta}_\alpha] =
		- \sum_{n=2}^\infty \eta_{\alpha,n}^\dagger \big(1 + \Delta_{\alpha,n}(\lambda)\big)\eta_{\alpha,n}~,
\end{align}
which for large $\lambda$ at leading order behaves as
\begin{align}
	\label{Seffll}
		S_{\alpha}^{\rm eff}[\boldsymbol{\eta}_\alpha] \,\underset{\lambda \to \infty}{\sim}\,
		 - \frac{4\pi^2}{\lambda \,s_\alpha} \sum_{n=2}^{\infty} n(n-1)\, \eta_{\alpha,n}^\dagger
		\eta_{\alpha,n}^{\phantom{\dagger}}~.
\end{align}
This expression represents a localized version of the effective action for the source fields
$\eta_{\alpha,n}(\vec{x})$ that are associated to the (normalized) operators 
$T_{\alpha,n}(\vec{x})$
of the quiver theory as discussed in Section~\ref{subsec:nh}. This action should therefore be connected to the on-shell value of the action (\ref{Seta}) upon enforcing the appropriate boundary conditions on the five-dimensional AdS fields $\eta_{\alpha,n}(x)$. We have not done this explicitly, but the quadratic form of (\ref{Seffll}) indeed matches; we will argue in Section~\ref{sec:discussion} that also the overall coefficient ${4\pi^2}/(\lambda \,s_\alpha)$ has a natural interpretation in this approach.

\subsection{Numerical checks}
\label{subsec:numerical}

We now present a few numerical checks of the analytic results obtained before. As in \cite{Beccaria:2021hvt}, we employ two independent methods.
The first one is a resummation  \`a la Pad\'e of the perturbative expansions of the quantities
$\Delta_{\alpha,n}(\lambda)$ that are obtained by inserting the definition (\ref{Xis}) of the $\Xx$ matrix into (\ref{gammaome}).
The second method relies on the fact that the supersymmetric localization provides an expression 
for the vacuum expectation value of a given observable in terms of a finite $N$-dimensional integral with a non-negative integrand (see for example (\ref{vevPz})). 
In particular, in the large $N$-limit, the integration domain shrinks around the saddle-point configuration so that the computation can be performed using Monte Carlo methods.

\subsubsection*{Pad\'e approximants}
As it was discussed in Section \ref{subsec:twistedcorr}, in the weak-coupling regime
the quantities $\Delta_{\alpha,n}(\lambda)$ are expressed as a perturbative series of the form
\begin{align}
\Delta_{\alpha,n}(\lambda) = \sum_{\ell=n}^{+\infty} c^{(\alpha,n)}_\ell
\Big(\frac{\lambda}{\pi^2}\Big)^\ell
\end{align}
where the coefficients $c^{(\alpha,n)}_\ell$ can be efficiently computed up to very high orders.
The first few of them can be read from the explicit formulas (\ref{explicitDelta}).
The radius of converge of these series is $\lambda=\pi^2$ (see \cite{Beccaria:2021hvt,Beccaria:2021vuc,Beccaria:2020hgy}). However, we can extract information on the 
region $\lambda > \pi^2$ by considering the series truncated at some order $L$ and
its diagonal Pad\'e approximant \cite{Baker}
\begin{align}
P_{[K/K]}(\Delta_{\alpha,n}) = \bigg[\,\sum_{\ell=n}^L c_\ell^{(\alpha,n)}
\Big(\frac{\lambda}{\pi^2}\Big)^\ell\,\bigg]_{[K/K]} \, \ ,
\end{align}
where $K$ is the degree of the polynomial which must satisfy $K < L/2$. The functions 
obtained in this way are shown for some cases by the solid lines in Figs.~\ref{fig:k2Z4} and \ref{fig:dim2} below. 

\subsubsection*{Monte Carlo methods}
In principle there are several Monte Carlo (MC) algorithms that can be used to evaluate the 
2-point functions (\ref{cTT}). Given the level of precision that we want to reach, we choose to employ a {Metropolis-Hastings algorithm} (see \cite{brooks2011handbook}), which was already considered in \cite{Beccaria:2021hvt,Beccaria:2021ksw}. We refer the reader to these works for technical details concerning its implementation, while here we just schematically summarize it. 
Given an initial configuration $Y$ of the eigenvalues $\{a_{I,u}\}$ (with $u=1,\ldots,N$) 
associated to the SU$(N)_I$ gauge group of the quiver matrix model, the algorithm generates a Markov chain of configurations $\{Y_n\}$ obeying detailed balance. Then the vacuum expectation value of an observable $\mathcal{O}(Y_j)$ is computed taking the arithmetic average over the elements of the chain, namely\footnote{Note that this vacuum expectation value is affected by both statistical and auto-correlation errors. Both these two sources of errors have been taken into account and estimated.}  
\begin{align}
\big\langle \mathcal{O}(Y_j) \big\rangle =  \frac{1}{n}\sum_{j=1}^{n}\mathcal{O}(Y_j) \, \ .
\end{align}
In principle MC methods can be applied for arbitrary values of the conformal dimension $n$ of the operators $\cT_{\alpha,n}$ and for arbitrary quiver theories. However, in order to be concrete and deal with an acceptable computational cost, here we have decided to focus on the cases $n=2,3$ for the quivers with $M=2,3,4$. We have performed the MC computations for different values of the pair 
$(N,\lambda)$ and observed that, at fixed $\lambda$ and for increasing values of $N$, 
the MC points tend to the Pad\'e curves. This is a clear sign of the validity of our numerical simulations.  

\subsection*{Results}
For $M=2$, we expect from (\ref{Deloddeven}) that
\begin{align}
  1 +  \Delta_{1,2}(\lambda)\,\underset{\lambda \to \infty}{\sim}\, \frac{8\pi^2}{\lambda} \qquad\mbox{and}\qquad
  1 +  \Delta_{1,3}(\lambda)\,\underset{\lambda \to \infty}{\sim}\, \frac{24\pi^2}{\lambda} ~,
\end{align}
and indeed this behavior is confirmed by the numerical checks we have performed. The analogous curves for the quivers with $M=3,4$ are simply obtained by multiplying the right-hand sides by
the factor $1/s_{\alpha}$ with $\alpha=1$, which from (\ref{salpha}) is
\begin{align}
\label{eq:predictions}
\frac{1}{s_1}=  \begin{cases} \frac{4}{3} \ \ \mathrm{for} \ \ M=3 ~, \\[1mm]
 2 \ \   \mathrm{for} \ \ M=4 ~.
\end{cases}
\end{align}
Again our numerical simulations in these cases confirm the expected behavior. 

\begin{figure}[ht]
    \centering
    \includegraphics[scale=0.35]{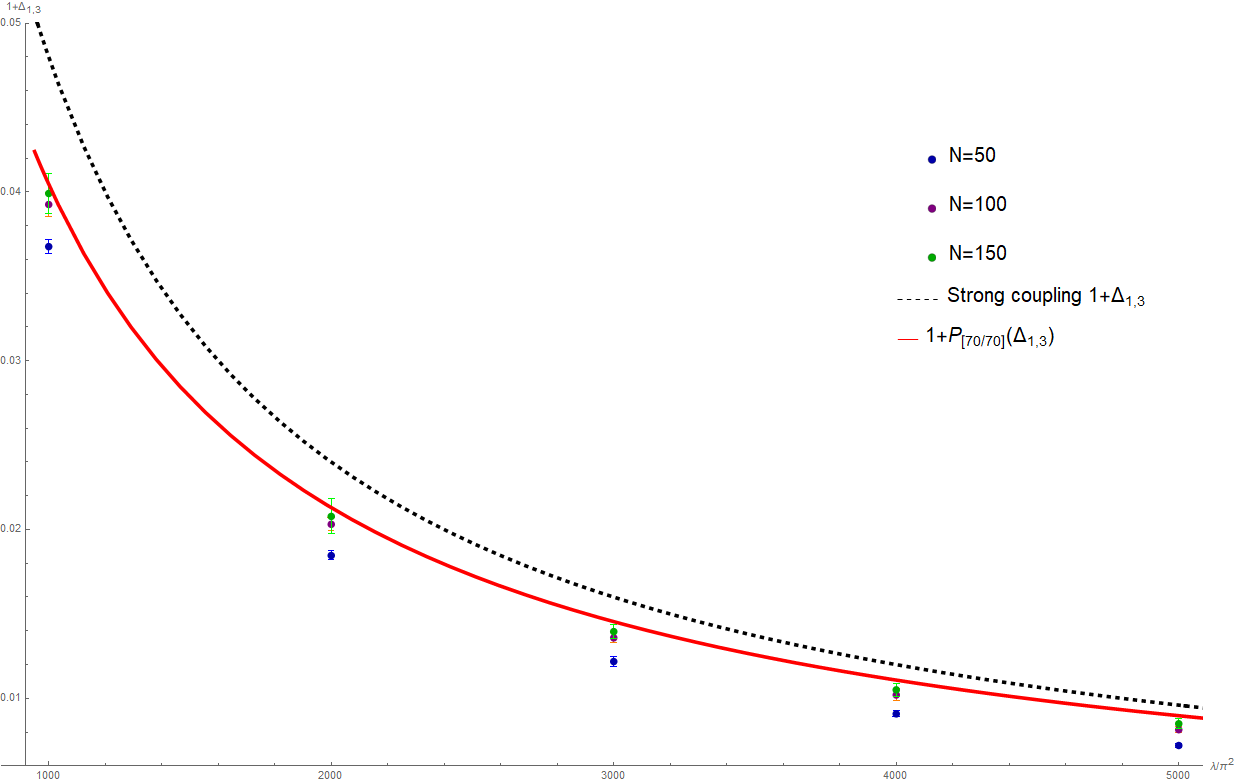}
    \caption{Pad\'e  curve  (red solid curve),  MC  data (colored points) and  the strong-coupling prediction (black dashed curve) for the function $1 + \Delta_{1,3}(\lambda)$ in the range $1000 \pi^2 \leq \lambda \leq 5000\pi^2$ of the $M=4$ quiver theory. The Pad\'e curve is obtained from the diagonal Pad\'e approximant of order 70 of the perturbative series, while the strong-coupling prediction is $48\pi^2/\lambda$. One clearly sees that, as $N$ increases, the MC points systematically tend towards the Pad\'e curve.}
    \label{fig:k2Z4}
\end{figure}

As an example, in Fig.~\ref{fig:k2Z4} we collect our results for the function 
$1+\Delta_{1,3}(\lambda)$ evaluated in the $M=4$ quiver theory with the methods discussed above.
The black dashed line represents the prediction based on the strong-coupling analysis, 
namely $48\pi^2/\lambda$; the red solid line represents the curve obtained from the Pad\'e approximant of order $K=70$ of the perturbative series. When $\lambda$ increases, 
the Pad\'e extrapolation nicely tends towards the strong-coupling curve. The colored points (with error bars) represent the results of the MC simulations for different values of $N$.
When $N$ increases we clearly see that, as anticipated, these points move towards 
the Pad\'e curve. We have observed the same features in all other cases we have considered.

\begin{figure}
    \centering
    \includegraphics[scale=0.38]{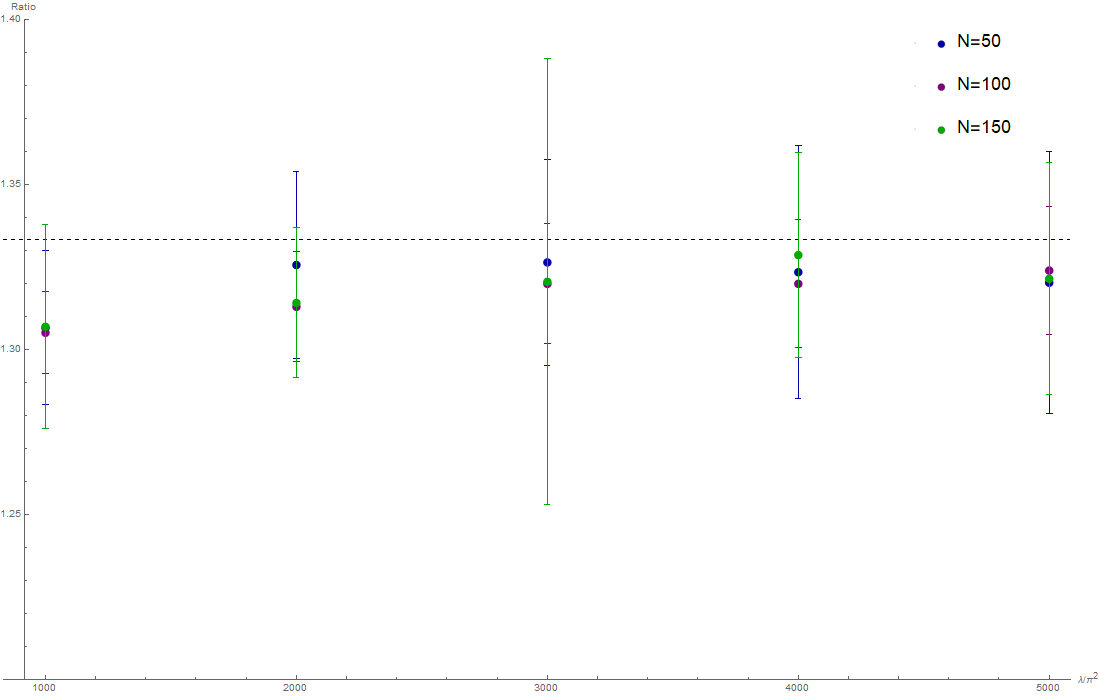}
    \caption{Ratio between the 2-point function \eqref{cTT} for $n=3$ in the circular quiver with $M=3$ and those found in the circular quiver with $M=2$, in the range $1000 \pi^2 \leq \lambda \leq 5000\pi^2$ and for $N=50,100,150$. The expected ratio is $4/3$.}
    \label{fig:ratio1}
\end{figure}

\begin{figure}
\centering
    \includegraphics[scale=0.39]{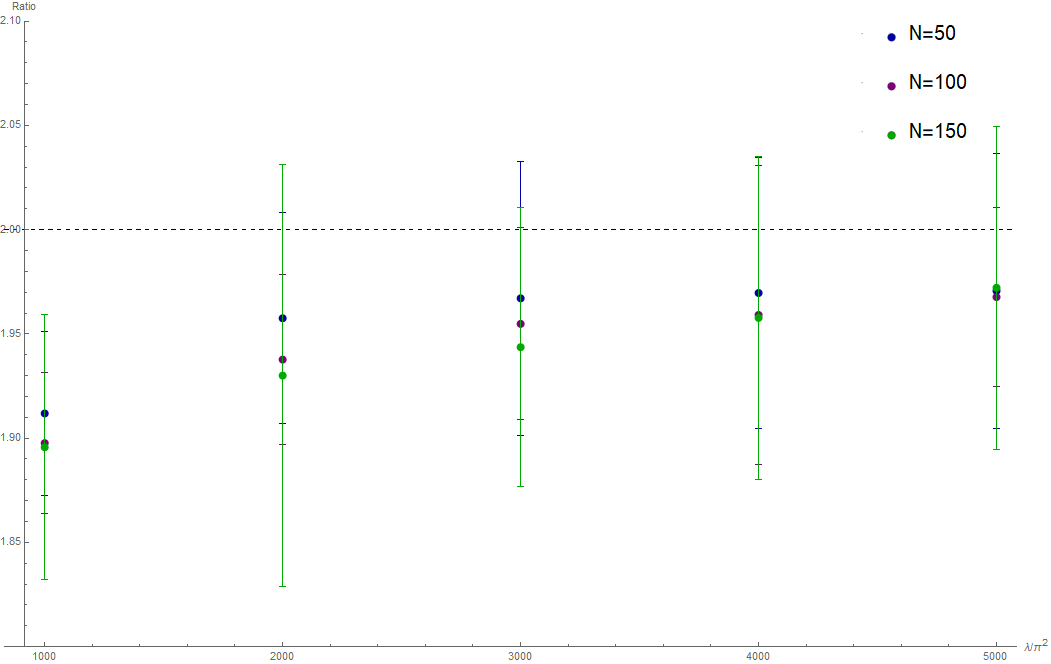}
    \caption{Ratio between the 2-point function \eqref{cTT} for $n=3$ in the circular quiver with $M=4$ and those found in the circular quiver with $M=2$, in the range $1000 \pi^2 \leq \lambda \leq 5000\pi^2$ and for $N=50,100,150$. The expected ration in this case is $2$.}
    \label{fig:ratio2}
\end{figure}

Moreover, in order to test the occurrence of the factors (\ref{eq:predictions}), 
we have fixed the values of $\lambda$ and, for different values of $N$, have evaluated numerically
the ratio between the 2-point functions (\ref{TTfin}) in the quivers with 3 and 4 nodes, respectively, and the same correlators in the quiver with 2 nodes. In Fig.~\ref{fig:ratio1} we report our findings
in the case $n=3$ for the ratio between the $M=3$ and $M=2$ quivers, which we expect to be $4/3$. Similarly, in Fig.~\ref{fig:ratio2} we report the results for the ratio between the $M=4$ and the $M=2$ theories, which we expect to be $2$. In both cases, taking into account the error bars, we observe that the MC points are indeed localized in a range very closed to the asymptotic theoretical prediction, and tend towards it for increasing values of $\lambda$.

\begin{figure}[ht]
    \centering
    \includegraphics[scale=0.35]{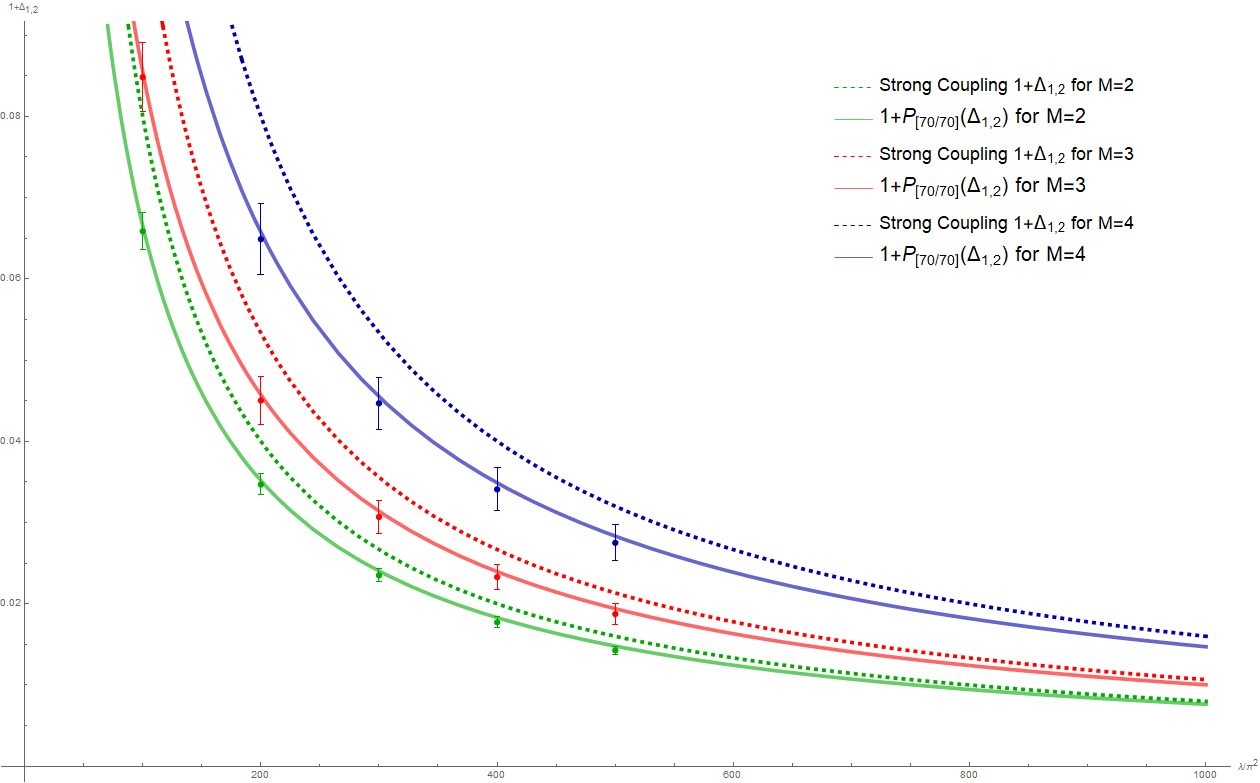}
    \caption{Pad\'e  curves (solid lines),  MC  data (points with error bars) and the large-$\lambda $ theoretical predictions for $1 + \Delta_{1,2}(\lambda)$ (dashed curves) in the range $100 \pi^2 \leq \lambda \leq 1000\pi^2$ for $N=50$ and $M=2$ (green), 3 (red) and 4 (blue).}
    \label{fig:dim2}
\end{figure}

Finally, we have considered the case $n=2$ in the quivers with $M=2,3,4$. 
Even if in principle there are no obstructions to use the same MC algorithm, in practice the computational cost in this case turns out to be higher. This mainly due the fact that the expected strong-coupling values for $n=2$ are much lower than those for operators 
of conformal dimension $n>2$. 
Therefore, in order to have an acceptable computational time and still be able to provide a good resolution between the MC points and the Pad\'e curve, we have considered a lower range for the coupling, namely $100 \leq \lambda/\pi^2 \leq 500$, and set $N=50$. The results we have obtained are
reported in Fig.~\ref{fig:dim2} where we observe that the MC points follow the behavior of the corresponding Pad\'e curves. Moreover, for large values of $\lambda$, the Pad\'e curves tend towards the corresponding strong-coupling theoretical predictions. We regard these features 
as a nice consistency check of the whole analysis.


\section{Discussion}
\label{sec:discussion}

The main result of this paper has been the calculation of the 2-point correlation
functions of the primary single-trace untwisted and twisted operators of 
the $\cN=2$ superconformal quiver theory in the large-$N$ limit. Normalizing to the 2-point functions of the $\cN=4$ SYM theory, we have found
\begin{subequations}
\begin{align}
	\frac{\big\langle U_{\alpha,n}^{\phantom{\dagger}}(\vec{x}) \,U_{\alpha,n}^{\,\dagger}(\vec{0})\big\rangle\phantom{\Big|}}{~\big\langle 
O_{n}(\vec{x})\, \overbar{O}_{n}(\vec{0})\big\rangle_{0}\phantom{\Big{|}}}&=
		1+O\big(N^{-2}\big)~,\label{res1}
		\\[1mm]
	\frac{\big\langle T_{\alpha,n}^{\phantom{\dagger}}(\vec{x}) \,T_{\alpha,n}^{\,\dagger}(\vec{0})\big\rangle\phantom{\Big|}}{~\big\langle 
O_{n}(\vec{x})\, \overbar{O}_{n}(\vec{0})\big\rangle_{0}\phantom{\Big{|}}}&=
		1+\Delta_{\alpha,n}(\lambda)	+O\big(N^{-2}\big)~,\label{res2}
\end{align}%
\end{subequations}
where the function $\Delta_{\alpha,n}(\lambda)$ is given in terms of the infinite matrix $\Xx$ as shown in (\ref{gammaome}). We stress the \emph{entire} dependence on $\lambda$ is captured in
this expression, so that we can use it to investigate the properties of the twisted correlators in the various regimes. When $\lambda$ is small, we can expand $\Delta_{\alpha,n}(\lambda)$ in power
series and obtain the perturbative expansion, retrieving at the first orders the results of the Feynman diagram calculations. When $\lambda$ is large, we exploit the properties of the Bessel functions contained in the $\Xx$ matrix and find that the leading term in the asymptotic expansion of
the twisted correlator is
\begin{equation}
\frac{\big\langle T_{\alpha,n}^{\phantom{\dagger}}(\vec{x}) \,T_{\alpha,n}^{\,\dagger}(\vec{0})\big\rangle\phantom{\Big|}}{~\big\langle 
O_{n}(\vec{x})\, \overbar{O}_{n}(\vec{0})\big\rangle_{0}\phantom{\Big{|}}}\,
\underset{\lambda \to \infty}{\sim} \,\,\frac{4\pi^2}{\lambda\,s_\alpha} n(n-1)~.
\label{res3}
\end{equation}
This result has been successfully checked against numerical simulations based on Monte Carlo methods.

It is interesting to observe that the $n$-independent prefactor in (\ref{res3}) has a nice interpretation in the holographic dual theory. Indeed, from the AdS/CFT correspondence \cite{Gubser:1998bc,Witten:1998qj} one knows that 
the 2-point functions of the conformal field theory are proportional to the
normalization of the supergravity action in the Anti-de Sitter background. For the $\cN=4$ SYM theory, the latter is the normalization of the supergravity action in ten dimensions, that is
\begin{equation}
\frac{1}{2\kappa_{10}^2}=\frac{1}{(2\pi)^7\,\alpha^{\prime\,4}\,g_s^2}=\frac{4N^2}{(2\pi)^5R^8}
\label{kappa10a}
\end{equation}
where the first equality follows from (\ref{kappa10}), while the last expression arises upon using
the fact that the radius $R$ of the Anti-de Sitter space is given by
\begin{equation}
R^4=4\pi g_s N\alpha^{\prime\,2}= \lambda\,\alpha^{\prime\,2}~.
\label{radius}
\end{equation}
On the other hand, the 2-point functions in the $\alpha$ twisted sector of the quiver theory are proportional to the normalization of the six-dimensional action of the scalars dual to the
twisted operators, which, as discussed in Section~\ref{subsec:nh} (see in particular (\ref{Seta})) is given by
\begin{equation}
\frac{1}{2\kappa_{6}^2}\,\frac{1}{\sin^2\big(\frac{\pi\alpha}{M}\big)}=\frac{4N^2}{(2\pi)^5R^4}
\,\frac{4\pi^2}{\lambda\,s_\alpha}
\label{kappa6a}
\end{equation}
where we have used (\ref{kappa6}), (\ref{salpha}) and (\ref{radius}). Thus, from this argument we
expect that the ratio of the twisted correlators in the quiver theory with respect to those of the $\cN=4$ SYM theory is proportional to the ratio of (\ref{kappa6a}) and (\ref{kappa10a}), which in units of the AdS radius is
\begin{equation}
\frac{4\pi^2}{\lambda\,s_\alpha}~.
\end{equation}
This is precisely what we find in (\ref{res3}) from the strong-coupling analysis of the matrix model results. Of course, as already mentioned, if one only considers the 2-point functions,
there is a normalization ambiguity in the holographic calculation \cite{Lee:1998bxa}. Nevertheless we find remarkable that our strong-coupling extrapolation of the matrix model results captures the $1/\lambda$ dependence suggested by the normalizations of the dual supergravity actions. It is therefore reasonable to expect that the remaining $n$-dependent factors in (\ref{res3}) should follow when the supergravity action for the normalized twisted scalars is localized on the AdS boundary by imposing appropriate boundary conditions.

It would be interesting to explicitly verify this fact and also to extend our analysis to higher-point correlation functions and to the sub-leading terms in the expansion for large $\lambda$ (some preliminary results on this have already been obtained \cite{desmet}). In this way one could investigate the properties of this asymptotic expansion, which is known to be non-Borel summable, and see what kind of non-perturbative completion is required to make it well-defined, similarly
to what has been recently discussed in \cite{Dorigoni:2021bvj,Dorigoni:2021guq} in the context to the $\cN=4$ SYM theory. In this respect, it would be interesting also to study the role played by instantons in the limit $N\to\infty$ with $g$ fixed both in the quiver models and in their orientifold descendants.

\vskip 1cm
\noindent {\large {\bf Acknowledgments}}
\vskip 0.2cm
We would like to thank Matteo Beccaria, Matthias Gaberdiel and Michelangelo Preti for useful discussions.
This research is partially supported by the INFN project ST\&FI
``String Theory \& Fundamental Interactions''. The work of F.G. is supported by a grant from the Swiss National Science Foundation, as well as via the NCCR SwissMAP.
The work of A.P. is supported by INFN with a``Borsa di studio post-doctoral per fisici teorici".
\vskip 1cm
\begin{appendix}
\section{Notations and conventions}
\label{app_notations}

Here we collect our notations and conventions for the various types of indices used throughout the
text:
\begin{itemize}
\item
labels of the $\mathbb{Z}_M$-quiver nodes: $I,J=0,\ldots ,M-1$,

\item
2-cycles of the $\mathbb{Z}_M$-orbifold singularity: $i,j,\ldots = 1,\ldots,M-1$,

\item
twisted sectors: $\alpha,\beta,\dots = 1,\ldots,M-1$,

\item
conformal dimensions: $n,m,\ldots=2,\ldots,\infty$

\item
SU$(N)$ adjoint indices: $a,b,c,\ldots = 1,\ldots,N^2-1$.

\item
SU$(N)$ fundamental indices: $u,v,\hat u,\hat v = 1,\ldots,N$.

\item
SU$(N)$ bi-fundamental indices: $A,B = 1,\ldots,N^2$.
\end{itemize}

\section{Properties of the matrix $\Xx^{\mathrm{even}}$ at strong coupling}
\label{app:b}

Consider the infinite matrix $\Xx^{\mathrm{even}}$ whose matrix elements are given in (\ref{Xevenis}).
To find its behavior for large $\lambda$, one can start by writing the products 
of Bessel functions appearing in its definition as an inverse Mellin transform, namely
\begin{align}
	\label{jj}
	J_{2k}\Big(\frac{t\sqrt{\lambda}}{2\pi}\Big) 
	J_{2\ell}\Big(\frac{t\sqrt{\lambda}}{2\pi}\Big) = 
	\int_{-\ii \infty}^{+\ii \infty}  \!\frac{ds}{2 \pi \ii}\,
	\frac{\Gamma(-s)\, \Gamma(2s+2k+2\ell+1)}
	{\Gamma(s+2k+1)\Gamma(s+2\ell+1)\Gamma(s+2k+2\ell+1)}
	\Big(\frac{t\sqrt{\lambda}}{4\pi}\Big)^{2s+2k+2\ell}~.
\end{align}
Inserting this expression in the definition (\ref{Xevenis}) and using the identity
\begin{align}
	\label{zet}
	\int_0^\infty \!dt\, \frac{\rme^t}{(\rme^t-1)^2}\, t^{2s+2k+2\ell-1} = 
	\Gamma (2s+2k+2\ell) \,\zeta(2s+2k+2\ell-1) \,
\end{align}
we can write the matrix elements of $\Xx^{\mathrm{even}}$ as follows:
\begin{align}
	\label{mtx}
	\mathsf{X}^{\mathrm{even}}_{k,\ell}
	=& - 8 (-1)^{k+\ell} \sqrt{2k\, 2\ell} \notag \\
	&\int_{-\ii \infty}^{+\ii \infty}\!  \frac{ds}{2 \pi {\rm i}}\,
	\frac{\Gamma(-s)\, \Gamma(2s+2k+2\ell+1) \Gamma(2s+2k+2\ell)\zeta(2s+2k+2\ell-1)}
	{\Gamma(s+2k+1)\Gamma(s+2\ell+1)\Gamma(s+2k+2\ell+1)}
	\Big(\frac{\sqrt{\lambda}}{4\pi}\Big)^{2s+2k+2\ell}\,.
\end{align}
When $\lambda\to \infty$, the asymptotic expansion of this expression receives contributions from the poles on the negative real axis and reads
\begin{align}
	\label{mtrx}
	\mathsf{X}^{\rm even}_{k,\ell}
	& = - \,8 (-1)^{k+\ell} \sqrt{2k\, 2\ell}\,
	\bigg[ \frac{\lambda}{16 \pi^2} \Big(\frac{\delta_{k-1,\ell}}{(2k-2)(2k-1)4k } +
	\frac{\delta_{k,\ell}}{(2k-1) 2k (2k+1)}
	\notag \\
	& \,\quad + \frac{\delta_{k+1,\ell}}{4k (2k+1)(2k+2)} 
	\Big) \,- \frac{\delta_{k,\ell}}{24 (2k)} \,+ O\Big(\frac{1}{\sqrt \lambda}\Big)\, \bigg]
	~.
\end{align}
This justifies the expression given in (\ref{xseven}) of the leading behavior 
of $\mathsf{X}^{\mathrm{even}}$ in terms of the matrix $\Sx^{\mathrm{even}}$ introduced in  (\ref{seven}).

To compute the invariants connected with $\mathsf{S}^{\mathrm{even}}$, 
it is convenient to introduce the asymmetric matrix $\mathsf{Y}$ with rational entries
\begin{align}
	\label{y}
	\mathsf{Y}_{k,\ell}=
	\Big(\frac{\delta_{k-1,\ell}}{4k (2k-1)} + \frac{\delta_{k,\ell}}{(2k-1)(2k+1)} +
	\frac{\delta_{k+1,\ell}}{4k(2k+1)}	\Big)~,
\end{align}
which is related to $\mathsf{S}^{\mathrm{even}}$ by a similarity transformation
\begin{equation}
	\mathsf{Y} =\mathsf{T}^{-1}\,\mathsf{S}^{\mathrm{even}}\,\mathsf{T}~,
	\label{sdy}
\end{equation}
where $\mathsf{T}$ is a diagonal matrix with entries
\begin{equation}
	\mathsf{T}_{k,\ell} = (-1)^k \, \sqrt{2k} \, \delta_{k,\ell}~.
	\label{dij}
\end{equation}
Of course, (\ref{sdy}) implies
$\det \mathsf{S}= \det \mathsf{Y}$ and $\tr \mathsf{S}= \tr \mathsf{Y}$.

In turn, $\mathsf{Y}$ can be written in terms of a matrix $\mathsf{\Lambda}$ with integer entries 
as follows:
\begin{equation}
	\mathsf{Y} =\mathsf{N}\,\mathsf{\Lambda}
	\label{yl}
\end{equation}
where
\begin{align}
	\mathsf{N}_{k,\ell}= \frac{\delta_{k\ell}}{2(2k-1)2k(2k+1)}~,
	\label{nn}
\end{align}
and
\begin{align}	
	\mathsf{\Lambda}_{k,\ell}= (2k+1)\,\delta_{k-1,\ell} + 4k\,\delta_{k,\ell} +
	(2k-1)\,\delta_{k+1,\ell}~.
	\label{lam}
\end{align}

In fact, we are interested in the determinants of the matrices $\Yx_{[k]}$ obtained from 
$\mathsf{Y}$ by removing the first $(k-1)$ rows and columns. In analogy with (\ref{yl}), we write
\begin{equation}
	\mathsf{Y}_{[k]} =\mathsf{N}_{[k]}\mathsf{\Lambda}_{[k]}~,
	\label{yln}
\end{equation}
so that
\begin{equation}
	\det \mathsf{Y}_{[k]} =
	\prod_{\ell=k}^{\infty} \Big(\frac{1}{4\ell(2\ell-1)(2\ell+1)}\Big)\, \det \mathsf{\Lambda}_{[k]}~.
	\label{detyn}
\end{equation}
The matrix $\mathsf{\Lambda}_{[k]}$ is explicitly given by
\begin{equation}
	\mathsf{\Lambda}_{[k]}=
	\begin{pmatrix}
		4k&2k-1&0&0&0&0& &&\\
		2k+3&4k+4&2k+1&0&0&0& &&\\
		0&2k+5&4k+8&2k+3&0&0&  & \cdots  &\\
		0&0&2k+7&4k+12&2k+5&0& &&\\
		&& \vdots &&&& &&\\
	\end{pmatrix}
	\label{matrlan}
\end{equation}
and its determinant is equal to the determinant of the matrix
\begin{equation}
 \mathsf{\Lambda'}_{[k]} =
	\begin{pmatrix}
		4k&2k-1&0&0&0&0& &&\\
		-2k+3&2k+5&2k+1&0&0&0& &&\\
		2k-3&0&2k+7&2k+3&0&0&  & \cdots  &\\
		-2k+3&0&0&2k+9&2k+5&0& &&\\
		&& \vdots &&&& &&\\
	\end{pmatrix}
	\label{matrlanp}
\end{equation}
which is obtained from $\mathsf{\Lambda}_{[k]}$ by summing with alternating signs its rows.
The determinant of $\mathsf{\Lambda'}_{[k]}$ can be easily computed by expanding with respect
to its first column. In this way we have
\begin{align}
	\label{detlp1}
	\det \mathsf{\Lambda'}_{[k]}
	& =   
	4k\, \prod_{\ell=k+2}^{\infty}(2\ell+1) +(2k-3) (2k-1) \prod_{\ell=k+3}^{\infty}(2\ell+1)
	\notag\\
	&\,\quad + (2k-3) (2k-1) (2k+1)\prod_{\ell=k+4}^{\infty}(2\ell+1)+ \ldots\\[1mm]
	& = \bigg[ \frac{4k}{2k+3}+ 
	\frac{(2k-3) (2k-1)}{(2k+3) (2k+5)}+ \frac{(2k-3) (2k-1)(2k+1)}{(2k+3) (2k+5)(2k+7) } + \ldots\bigg]\prod_{\ell=k+1}^{\infty}(2\ell+1)~.\notag
\end{align}	
With a few simple manipulations we can rewrite the last line and get
\begin{align}
	\det \mathsf{\Lambda'}_{[k]}
	& =   
	\bigg[ \frac{4k}{2k+3}-1 
	-\frac{k-3/2}{k+3/2}+ \sum_{n=0}^{\infty}
	\frac{(k-3/2)_n}{(k+3/2)_n}\bigg] \prod_{\ell=k+1}^{\infty}(2\ell+1)\notag\\
	&= 	\sum_{n=0}^{\infty}
	\frac{(k-3/2)_n}{(k+3/2)_n}\,\prod_{\ell=k+1}^{\infty}(2\ell+1)
=F(k-3/2,1,k+3/2;1)\prod_{\ell=k+1}^{\infty}(2\ell+1)~.
	\end{align}
where $(a)_n$ is the Pochhammer symbol and $F$ is the hypergeometric function. From the properties of the latter, we obtain
	\begin{align}
	\det \mathsf{\Lambda'}_{[k]}
	&= \frac{\Gamma (k+3/2)\, \Gamma(2)}{\Gamma(3) \, \Gamma(k+1/2)} \,\prod_{\ell=k+1}^{\infty}(2\ell+1)	\notag \\
	&=\frac{2k+1}{4} \prod_{\ell=k+1}^{\infty}(2\ell+1) = 
	\frac{1}{4} \prod_{\ell=k}^{\infty}(2\ell+1)~.
		\label{detlp3}
\end{align}
Using (\ref{detlp3}) in (\ref{detyn}), we finally have
\begin{equation}
	\det \mathsf{Y}_{[k]} =\frac{1}{4} \prod_{\ell=k}^{\infty} 
	\Big(\frac{1}{4\ell(2\ell-1)}\Big) ~.
	\label{detynr}
\end{equation}
As discussed in Section~\ref{subsec:sc}, to determine the strong coupling behavior of the 2-point functions of twisted operators of dimension $2k$, we need to compute the ratio $\det S^{\mathrm{even}}_{[k+1]}/\det S^{\mathrm{even}}_{[k]}$.
Taking into account the observation after (\ref{dij}) and using the result (\ref{detynr}), we find
\begin{equation}
	\frac{\det \mathsf{S}^{\mathrm{even}}_{[k+1]}\phantom{\Big|}}{\det \mathsf{S}^{\mathrm{even}}_{[k]}\phantom{\Big|}}
	=\frac{\det \mathsf{Y}_{[k+1]}}{\det \mathsf{Y}_{[k]}} 
	= 4k \,(2k-1)~,
	\label{yn11}
\end{equation}
which is the formula in (\ref{Seven11}) of the main text.

\section{The $M=2$ quiver theory and its orientifold}
\label{app:Z2}

\begin{figure}[ht]
\begin{center}
\includegraphics[scale=0.8]{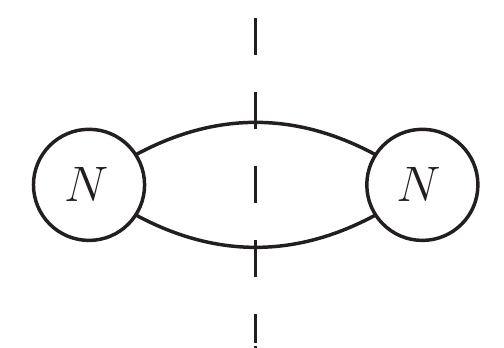}
\end{center}
\caption{The quiver diagram for the $M=2$ theory; the dashed line represents the mirror effect of the orientifold action.}
\label{Fig:quiver2nodes}
\end{figure}

In this appendix we consider in detail the quiver theory with $2$ nodes represented in
Fig.~\ref{Fig:quiver2nodes}.

This quiver theory is particularly
interesting since it is the parent of the $\cN=2$ superconformal SU($N$) gauge theory with one symmetric and one anti-symmetric hypermultiplet that arises by
taking a suitable orientifold projection (see for example \cite{Park:1998zh}). This latter theory, 
also called $\mathbf{E}$ theory in \cite{Billo:2019fbi}, has been recently studied in detail in
a series of papers \cite{Beccaria:2020hgy,Beccaria:2021vuc,Beccaria:2021hvt} using matrix model techniques, and some of its strong coupling properties have been elucidated. Moreover, this orientifold theory admits an holographic dual of the form AdS$_5\times S^5/\mathbb{Z}_2$ \cite{Ennes:2000fu}. We now provide some details on the connection between the 2-node quiver
theory and the $\mathbf{E}$ theory.

When $M=2$, there is only one self-conjugate twisted sector and thus the index $\alpha$ used in the main text takes only one value and
can be suppressed. In this quiver theory, the single-trace chiral primary operators,
written in terms of the chiral fields of the two vector multiplets, are
\begin{subequations}
\begin{align}
U_n(\vec{x})&=\frac{1}{\sqrt{2}}\Big(\tr \Phi_0(\vec{x})^n+\tr \Phi_1(\vec{x})^n\Big)~,\label{Un2}\\
T_{n}(\vec{x})&=\frac{1}{\sqrt{2}}\Big(\tr \Phi_0(\vec{x})^n-\tr \Phi_1(\vec{x})^n\Big)~.\label{Tnalpha2}
\end{align}
\label{UT2}%
\end{subequations}
The anti-chiral operators $\overbar{U}_n$ and $\overbar{T}_n$ are defined in a similar manner with the chiral fields replaced by their complex conjugate, so that (see (\ref{conjugation}))
\begin{equation}
\overbar{U}_n=U_n^\dagger\qquad\mbox{and}\qquad
\overbar{T}_n=T_n^\dagger~.
\end{equation}
We are interested in the 2-point functions of these operators, normalized with respect to those of the
$\cN=4$ SYM theory, namely
\begin{align}
	\label{cTT2}
	\frac{\big\langle U_{n}(\vec{x}) \,\overbar{U}_{n}(\vec{0})\big\rangle\phantom{\Big|}}{~\big\langle 
O_{n}(\vec{x})\, \overbar{O}_{n}(\vec{0})\big\rangle_{0}\phantom{\Big{|}}}\qquad\mbox{and}\qquad
	\frac{\big\langle T_{n}(\vec{x}) \,\overbar{T}_{n}(\vec{0})\big\rangle\phantom{\Big|}}{~\big\langle 
O_{n}(\vec{x})\, \overbar{O}_{n}(\vec{0})\big\rangle_{0}\phantom{\Big{|}}}~.
\end{align}
which we compute using the matrix model techniques explained in the main text.

{From} the formulas in Section~\ref{sec:matrix}, we see that the partition function of the matrix model corresponding to the $M=2$ quiver theory is
\begin{align}
\label{intda12}
		\cZ &= \int da_0\,da_1~ \rme^{-\tr a_0^2-\tr a_1^2~ -S_{\mathrm{int}}}
\end{align}
where
\begin{align}
		S_{\mathrm{int}} & = 2\sum_{m=2}^\infty \sum_{k=2}^{2m}
		(-1)^{m+k} \Big(\frac{g^2}{8\pi^2}\Big)^{\!m} \,\binom{2m}{k}
		\,\frac{\zeta(2m-1)}{2m} 
		\big(\tr a_{0}^{2m-k} - \tr a_{1}^{2m-k}\,\big) \big(\tr a_{0}^{k} - \tr a_{1}^{k}\big)
		~.
\end{align}
It is manifest from this expression that only the twisted combinations
\begin{align}
\big(\tr a_0^{\ell} -\tr a_1^{\ell}\big)
\end{align}
appear in the interaction action $S_{\mathrm{int}}$. 
Such combinations are not orthogonal to each other, not even in the
free Gaussian model. Performing the Gram-Schmidt procedure, we then introduce the 
normal-ordered twisted combinations $\cP_\ell$, which are orthonormal with respect to the
Gaussian measure (see (\ref{canT})), and rewrite $S_{\mathrm{int}}$ as
\begin{align}
	\label{keySZ2}	
S_{\mathrm{int}} &= - \frac{1}{2} \sum_{k,\ell=2}^\infty \cP_k \,\Xx_{k,\ell}\, \cP_\ell\notag\\[1mm]
&=- \frac{1}{2} \sum_{k,\ell=1}^\infty \cP_{2k} \,\Xx^{\mathrm{even}}_{k,\ell}\, \cP_{2\ell}	
- \frac{1}{2} \sum_{k,\ell=1}^\infty \cP_{2k+1} \,\Xx^{\mathrm{odd}}_{k,\ell}\, \cP_{2\ell+1}
\end{align}
where the infinite matrices $\Xx$, $\Xx^{\mathrm{even}}$ and $\Xx^{\mathrm{odd}}$ are
defined in \eqref{Xeo0} -- \eqref{Xevenis} of the main text. Then, using these expression and
following the general procedure explained in Sections~\ref{subsecn:gaussian} and \ref{subsec:twistedcorr}, in the large-$N$ limit we find
\begin{subequations}
\begin{align}
\frac{\big\langle U_{n}(\vec{x}) \,\overbar{U}_{n}(\vec{0})\big\rangle\phantom{\Big|}}{~\big\langle 
O_{n}(\vec{x})\, \overbar{O}_{n}(\vec{0})\big\rangle_{0}\phantom{\Big{|}}} &=~1+O\big(N^{-2}\big)~,\label{untwZ2}\\
\frac{\big\langle T_{n}(\vec{x}) \,\overbar{T}_{n}(\vec{0})\big\rangle\phantom{\Big|}}{~\big\langle 
O_{n}(\vec{x})\, \overbar{O}_{n}(\vec{0})\big\rangle_{0}\phantom{\Big{|}}}&= ~1+\Delta_n(\lambda)+O\big(N^{-2}\big)~,
\label{twZ2}
\end{align}
\end{subequations}
where
\begin{align}
	\label{gammaomeZ2}
		1 + \Delta_{2k}(\lambda) =
		\bigg(\frac{1}{\mathbb{1} - \Xx^{\mathrm{even}}_{[k]}}\bigg)_{1,1} \qquad\mbox{and}\qquad
		1 + \Delta_{2k+1}(\lambda) =
		\bigg(\frac{1}{\mathbb{1} - \Xx^{\mathrm{odd}}_{[k]}}\bigg)_{1,1} ~,
\end{align}
in agreement with the general formulas (\ref{gammaome}).

We now analyze what happens when an orientifold projection is performed and the $M=2$ quiver
theory is reduced to the $\mathbf{E}$ theory. This orientifold action enforces some identifications of some states of the original quiver theory, which are obtained by means of 
\begin{equation}
\Phi_0(\vec{x})\to \Phi(\vec{x})~, \qquad \Phi_1(\vec{x})\to -\Phi(\vec{x})~,
\label{identification}
\end{equation}
where $\Phi$ is the scalar field of the single vector multiplet of the $\mathbf{E}$ theory.
As a consequence of this identification, only a subset of the untwisted and twisted operators (\ref{UT2}) survive. Specifically, only the untwisted operators of even dimension and the
twisted operators of odd dimensions are kept, while all others are removed. Indeed, we have
\begin{equation}
\begin{cases}
U_{2k}(\vec{x})~\to~ \sqrt{2} \,\tr \Phi^{2k}(\vec{x})\,\equiv\, \sqrt{2}\,\,O_{2k}(\vec{x})\\[2mm]
U_{2k+1}(\vec{x})~\to~0\\[2mm]
T_{2k}(\vec{x})~\to~0\\[2mm]
T_{2k+1}(\vec{x})~\to~ \sqrt{2} \,\tr \Phi^{2k+1}(\vec{x})\,\equiv\, \sqrt{2}\,\,O_{2k+1}(\vec{x})
\end{cases}
\end{equation}
We therefore see that in the $\mathbf{E}$ theory the chiral operators $O_{2k}$ of even dimension arise from untwisted combinations in the original quiver theory, while the chiral operators
$O_{2k+1}$ of odd dimension are of twisted type. This fact was already pointed out in
\cite{Beccaria:2021hvt} and has a nice counterpart in the holographically dual description as observed
in \cite{Ennes:2000fu}.

In the matrix model description the identification (\ref{identification}) is implemented by the rule
\begin{equation}
a_0\to a~, \qquad a_1\to -a~.
\label{identOrientifold}
\end{equation}
Correspondingly, the partition function (\ref{intda12}) becomes
\begin{equation}
\cZ^{\,\mathbf{(E)}}= \int\!da~ \rme^{-\tr a^2~ -S_{\mathrm{int}}^{\,\mathbf{(E)}}}
\end{equation}
where\,%
\footnote{The interaction actions $S_{\mathrm{int}}$ and $S_{\mathrm{int}}^{\,\mathbf{(E)}}$
are related as follows: $S_{\mathrm{int}}^{\,\mathbf{(E)}}=\frac{1}{2}\,S_{\mathrm{int}}\Big|_{a_0\,=\,-a1\,=\,a}$.}
\begin{align}
S_{\mathrm{int}}^{\,\mathbf{(E)}}&=4 \!\sum_{\ell,m=1}^\infty(-1)^{m+\ell}
\Big(\frac{g^2}{8\pi^2}\Big)^{m+\ell+1}
\frac{(2m+2\ell+1)!}{(2m+1)!(2\ell+1)!}\,\zeta(2m+2\ell+1) \,\tr a^{2m+1}\,\tr a^{2\ell+1} ~.
\end{align}
Only the traces of odd powers of $a$ appear in the interaction action, which therefore
can be written as 
\begin{align}
	\label{SintOrient}	
S_{\mathrm{int}}^{\,\mathbf{(E)}}= - \frac{1}{2} \sum_{k,\ell=1}^\infty \cP_{2k+1} \,\Xx^{\mathrm{odd}}_{k,\ell}\, \cP_{2\ell+1}	
\end{align}
where $\cP_{2k+1}$ are the orthonormal combinations obtained from $\tr a^{2k+1}$ by applying the Gram-Schmidt procedure with respect to the Gaussian measure. This is precisely the
same expression obtained in \cite{Beccaria:2021hvt}.

Using this matrix model and following the procedure explained above, it is immediate to obtain the 2-point functions of the primary operators in the $\mathbf{E}$ theory. For those with even dimension, which
stem from untwisted operators of the quiver theory, we simply 
have to use (\ref{untwZ2}), while for the operators of odd dimension, which are of twisted type, we
read the result from (\ref{twZ2}). Explicitly, we have
\begin{subequations}
\begin{align}
\frac{\big\langle O_{2k}(\vec{x}) \,\overbar{O}_{2k}(\vec{0})\big\rangle\phantom{\Big|}}{~\big\langle 
O_{2k}(\vec{x})\, \overbar{O}_{2k}(\vec{0})\big\rangle_{0}\phantom{\Big{|}}} &=~1+O\big(N^{-2}\big)~,\\
\frac{\big\langle O_{2k+1}(\vec{x}) \,\overbar{O}_{2k+1}(\vec{0})\big\rangle\phantom{\Big|}}{~\big\langle 
O_{2k+1}(\vec{x})\, \overbar{O}_{2k+1}(\vec{0})\big\rangle_{0}\phantom{\Big{|}}}&= ~1+\Delta_{2k+1}(\lambda)+O\big(N^{-2}\big)~,
\end{align}
\end{subequations}
where $\Delta_{2k+1}(\lambda)$ is defined in (\ref{gammaomeZ2}). These results are in full agreement with \cite{Beccaria:2021hvt}.

\end{appendix}


\providecommand{\href}[2]{#2}\begingroup\raggedright\endgroup

\end{document}